\documentclass{aa}

\usepackage{tensor}
\usepackage{graphicx}   
\usepackage{amsmath}    
\usepackage{amssymb}    

\usepackage[symbol]{footmisc}

\usepackage{booktabs}   
\usepackage{color,bm}
\usepackage[colorlinks=true, linkcolor = blue, anchorcolor = blue, citecolor = blue, filecolor = blue, urlcolor = blue]{hyperref}
\usepackage{comment}

\defcitealias{brinkerink15}{BR15}
\defcitealias{brinkerink21}{BR21}

\graphicspath{{./}{Figures/}}

\title{Magnetic flux eruptions at the root of time lags in low-luminosity active galactic nuclei}
\titlerunning{Magnetic flux eruptions at the root of time lags in low-luminosity AGN}

\author{Jesse Vos\inst{\ref{inst1}}\thanks{\email{jt.vos@astro.ru.nl}} \and Jordy Davelaar\inst{\ref{inst2},\ref{inst3}} \and Hector Olivares\inst{\ref{inst1},\ref{inst4}} \and Christiaan Brinkerink\inst{\ref{inst1}} \and Heino Falcke\inst{\ref{inst1}}}
\authorrunning{Vos et al.}

\date{\today}

\institute{
Department of Astrophysics/IMAPP, Radboud University, PO Box 9010, 6500 GL Nijmegen, the Netherlands\label{inst1} \and
Center for Computational Astrophysics, Flatiron Institute, 162 Fifth Avenue, New York, NY 10010, USA\label{inst2} \and
Department of Astronomy and Columbia Astrophysics Laboratory, Columbia University, 550 W 120th St, New York, NY 10027, USA\label{inst3} \and
Departamento de Matem\'{a}tica da Universidade de Aveiro and Centre for Research and Development in Mathematics and Applications (CIDMA), Campus de Santiago, 3810-193 Aveiro, Portugal \label{inst4}
}


\abstract {Sagittarius A$^\ast$ is a compact radio source at the center of the Milky Way that has not conclusively shown evidence to support the presence of a relativistic jet. Nevertheless, indirect methods at radio frequencies do indicate consistent outflow signatures. } 
{Temporal shifts between features in frequency bands are known as time lags, associated with flares or outflows of the accretion system. 
It is possible to gain information on the emission and outflow mechanics by interpreting these time lags.
}
{By means of a combined general-relativistic magnetrohydrodynamical and radiative transfer modeling approach, we studied the origin of the time lags for magnetically arrested disk models with three black hole spins ($a_\ast \in \{ -0.9375, 0, 0.9375 \}$).
We exclusively modeled the emission from the source across a frequency range of $\nu=19-47$ GHz.
Our study also includes a targeted ``slow light'' investigation for one of the best-fitting ``fast light'' windows.
}
{
We were able to recover the observational time-lag relations in various windows of our simulated light curves.
The theoretical interpretation of these most promising time-lag windows is threefold: i) a magnetic flux eruption perturbs the jet-disk boundary and creates a flux tube; ii) the flux tube orbits and creates a clear emission feature; and iii) the flux tube interacts with the jet-disk boundary.
The best-fitting windows have an intermediate (i=30$^\circ$/50$^\circ$) inclination and zero black hole spin.
The targeted slow light study did not produce better-fitting time lag results, which indicates that the fast versus slow light paradigm is often not intuitively understood and is likely to be influential in timing-sensitive black hole accretion studies.  
}
{While previous studies have sought to interpret time-lag properties with spherical or jetted expansion models, we show that this picture is too simplistic. Sophisticated general-relativistic magnetrohydrodynamical models consistently capture the observational time-lag behavior, which is rooted in the complex dynamic interplay between the flux tube and coupled disk-jet system. 
}

\newcommand{\sgr}{Sgr A$^\ast$\xspace}
\newcommand{\MP}{{\tt MADP}\xspace}
\newcommand{\MZ}{{\tt MAD0}\xspace}
\newcommand{\MM}{{\tt MADM}\xspace}
\newcommand{\rg}{$r_\mathrm{g}$\xspace}
\newcommand{\rgc}{$r_\mathrm{g}/c$\xspace}

\begin{document}

\maketitle 

\section{Introduction} \label{int:introduction}

Relativistic jets have been established as a prominent feature of some active galactic nuclei (AGN).
Even though prominent jets are present for M87 \citep{owen89,degasperin12} and Centaurus A \citep{israel98,hardcastle03,janssen21}, it is not a common feature of AGN in general \citep{hada20}.
The compact radio source Sagittarius A* (Sgr A*) at the center of the Milky Way does not boast a direct jet detection, even though claims of jet-like features spanning several parsecs have been made \citep{li13}. The exact origin of Sgr A$^\ast$'s emission signature is still highly debated \citep[as described in, e.g.,][]{issaoun19,eht22sgrav}, but we can infer the presence of a (compact or unresolved) jet from the flat or inverted radio spectrum that is difficult to create without an extended emission structure.

Supermassive black holes (SMBHs) have been pivotal in understanding the origin and dynamics of jets \citep{blandford19}.
Collimated outflows from SMBH systems are expected as a means of angular momentum and energy dissipation either via the Blandford-Znajek \citep[BZ;][]{blandford77} or Blandford-Payne \citep[BP;][]{blandford82} processes.
While the former invokes the rotation of the central BH as the main driver, the latter relies on the rotational energy of the disk itself to produce outflows or winds. 
If sufficient poloidal magnetic field is present and the BH has a non-negligible spin, then the BZ mechanism is assumed to be dominant \citep[e.g.,][]{tchekhovskoy12b}.

\citealt{brinkerink15,brinkerink21} (hereafter \citetalias{brinkerink15,brinkerink21}) observed flares at various radio frequencies ($\nu=19-47$ GHz) and quantified the differences in the arrival times of light curve features as a function of frequency with a cross-correlation analysis.
The acquired ``time lags'' can be used to infer properties of the emitting plasma, mostly with regard to the flow velocity.
Previously, time lags have been interpreted with simple in- or outflow models, either via a jet-dominated \citep{falcke09} or an expanding cloud model \citep{vanderlaan66,yusef-zadeh08}.
Until now, however, no rigorous study based on general relativistic magnetohydrodynamical (GRMHD) simulations of BH accretion had been undertaken in 3D (some preliminary work was done in 2D by \citealt{okuda23}).
Such GRMHD simulations describe the global accretion dynamics around the central compact object and are known to self-consistently bring about a coupled disk-jet system with the associated in- and outflows.  

In this paper, we investigate whether GRMHD models of Sgr A$^\ast$ can self-consistently capture the time lags, as seen by \citetalias{brinkerink15}.
Our modeling effort starts from a new set of GRMHD simulations with various black hole spin parameters that are post-processed to acquire synchrotron emission images and light curves.
Due to its low Eddington luminosity, accretion onto Sgr A$^\ast$ is best described by an advection-dominated and radiatively inefficient accretion flow \citep[ADAF and RIAF;][]{narayan94,narayan95,abramowicz95}. 
Coupled jet-ADAF models were subsequently developed to better recover the spectra of Sgr A$^\ast$, and low-luminosity AGN in general \citep{yuan02}.
Dynamically evolved GRMHD models, which naturally display all previously mentioned characteristics, are typically split in two subclasses: SANE \citep[from Standard And Normal Evolution; e.g.,][]{devilliers03} or MAD \citep[from Magnetically Arrested Disk; e.g.,][]{igumenshchev03,narayan03}.
While the former is relatively weakly magnetized and turbulence-driven, the latter has a larger, more highly magnetized disk with a coherent large-scale magnetic field.

For this work, we limit ourselves to the investigation of MAD models since they are currently favored for Sgr A$^\ast$ \citep{eht22sgrav,eht22sgrai}.
A MAD model characteristic is that they display magnetic flux eruptions, associated with a critical accumulation of magnetic field, that produce under-dense and over-magnetized regions in the accretion disk called flux tubes \citep{mckinney12,dexter20,porth21}.
These flux tubes are the end product of magnetic reconnection between the upper and lower hemispheres that culminates in a vertical, poloidal field structure \citep{ripperda20,ripperda22,davelaar23}. 
Flux tubes turn out to create clear synchrotron emission features \citep[also shown in][]{mahdi23,davelaar23} and are prominently represented in the best-corresponding simulation time-lag windows at the evaluated radio frequencies ($\nu = 19 - 47$ GHz).
Moreover, we find evidence for interaction between the flux tube exhaust and jet-disk boundary, which is a prominent source of variability in the light curves.

The paper is structured as follows.
The methods include a brief overview of our GRMHD simulation configuration, the post-processing procedure, and how one calculates the time lags. This is all outlined in Sect.~\ref{meth:methods}.
The results and their interpretation are presented in Sect.~\ref{res:results}.
The discussion and conclusion can be found in Sect.~\ref{sec:discconc}.

\section{Methods} \label{meth:methods}

In the following sections, we will outline the methods employed for our GRMHD simulations (Sect.~\ref{meth:grmhd}), the radiative transfer procedure (Sect.~\ref{meth:grrt}), the cross-correlation procedure and observational goodness-of-fit estimation (Sect.~\ref{meth:lccf}), and an overview of our slow light implementation within the radiative transfer method (Sect.~\ref{meth:slow_light}).

\begin{figure*}
    \centering
    \includegraphics[width=\textwidth]{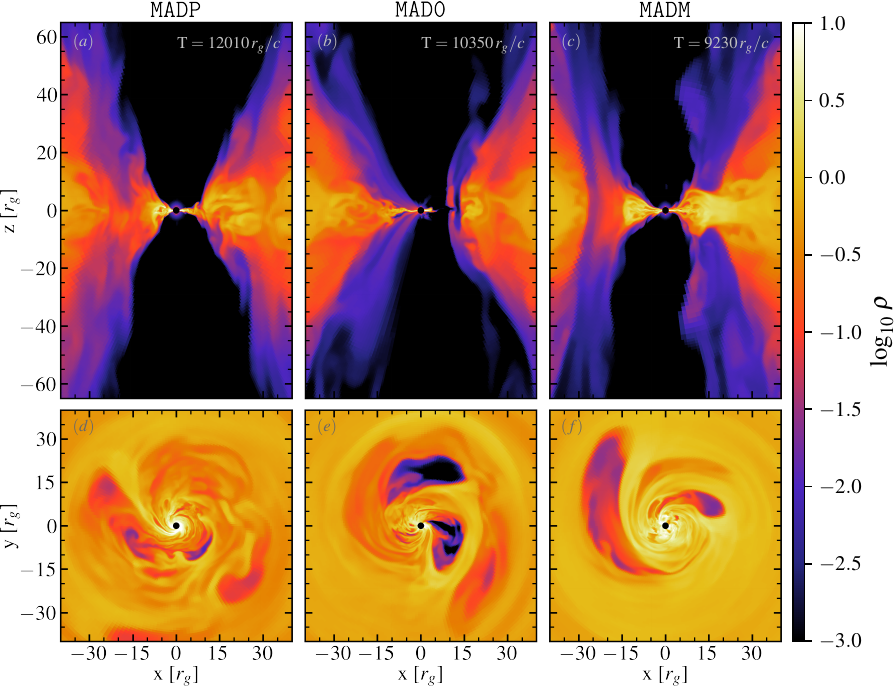}
    \caption{Flux tubes in the 3D GRMHD simulations. They are visualized with meridial (panels $a$, $b$, $c$) and equatorial ($d$, $e$, $f$) slices and correspond to low-density ($\rho$), high-magnetization ($\sigma$ in Fig. \ref{fig:gridtempapp}) regions that come about after a magnetically driven flux eruption starting from the central BH. The chosen slices correspond to promising time-lag windows, as listed in Table \ref{tab:chi2}. Most jet-related emission is expected from low-density ($\rho \sim 10^{-2}$) transition region known as the jet sheath (see panels $a$, $b$, $c$). }
    \label{fig:3DGRMHD}
\end{figure*}

\begin{figure}[!h]
    \centering
    \includegraphics[width=\columnwidth]{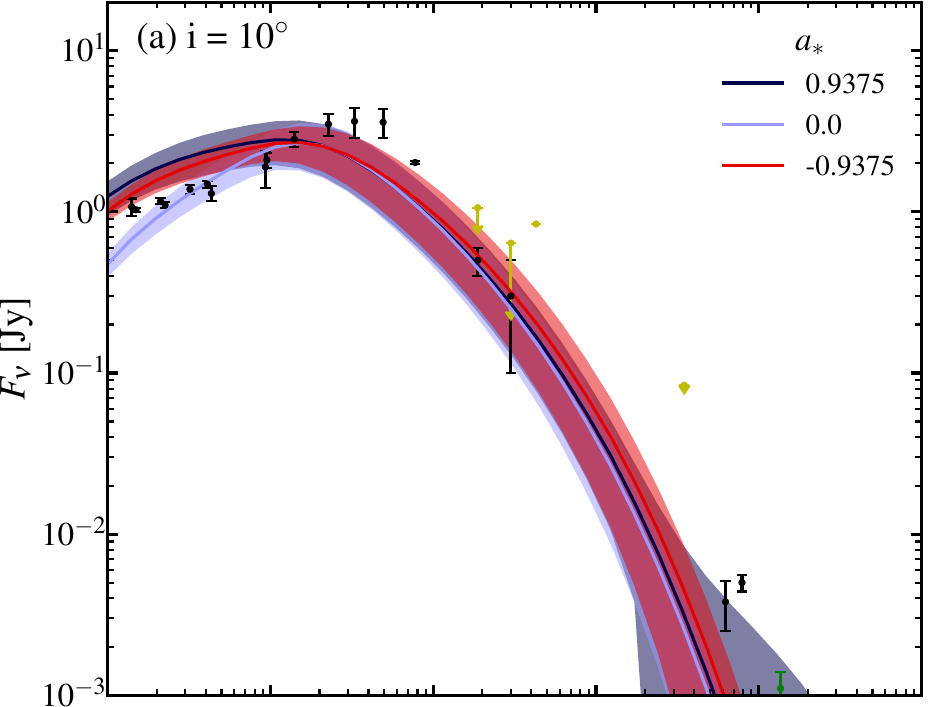}
    \includegraphics[width=\columnwidth]{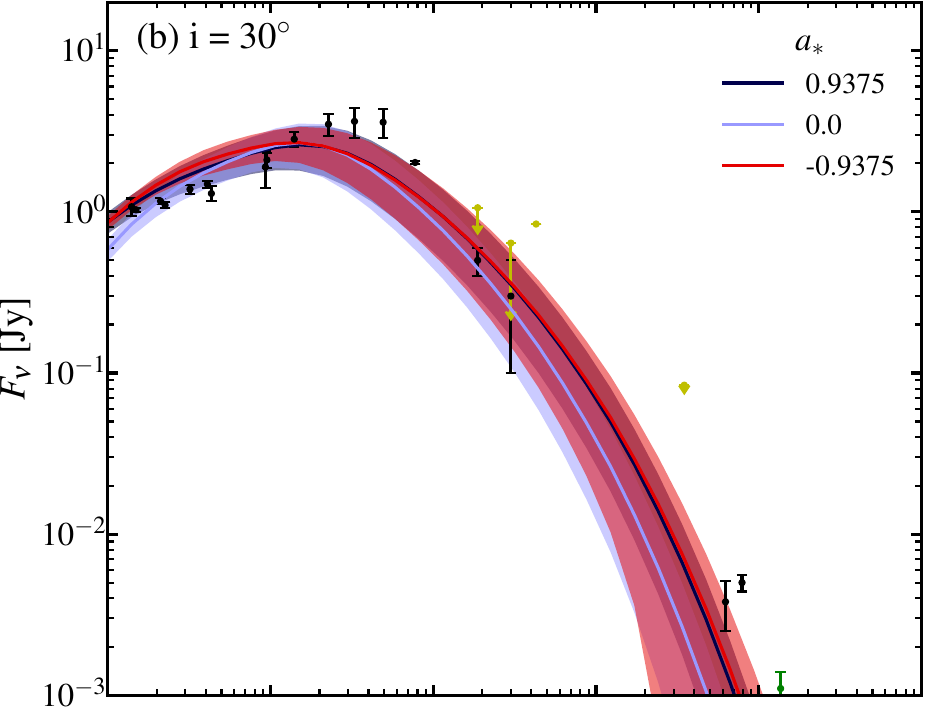}
    \includegraphics[width=\columnwidth]{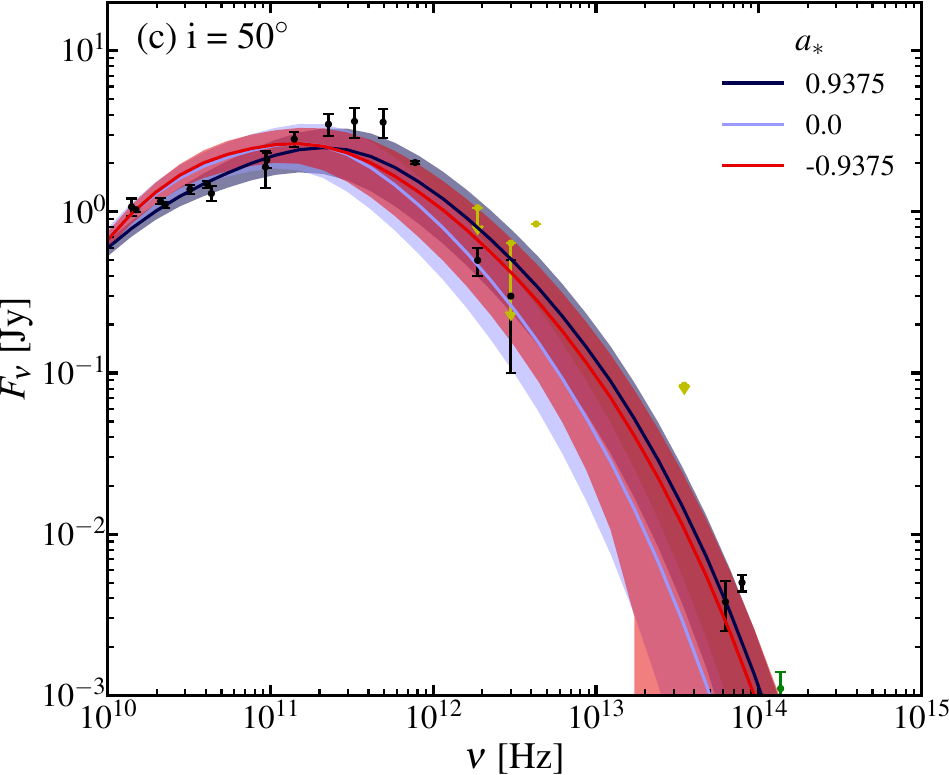}
    \caption{SEDs for three inclinations. From top to bottom, we find $i=10^\circ$ (panel a), $i=30^\circ$ (b), and $i=50^\circ$ (c). Each panel contains the fit corresponding to the models \MP ($a_\ast = 0.9375$ in \emph{dark blue}), \MZ ($a_\ast = 0.0$ in \emph{cyan}), and \MM ($a_\ast = -0.9375$ in \emph{red}), where the shaded region denotes the one-sigma standard deviation for the $1$ \rgc spaced light curves per frequency. The origin of the observational points are listed in Sect.~\ref{res:SED}. }
    \label{fig:SED}
\end{figure}

\subsection{General relativistic magnetohydrodynamics} \label{meth:grmhd}
We simulate the accretion flow surrounding a rotating Kerr black hole with the Black Hole Accretion Code \citep[{\tt BHAC},][]{porth17,olivares19}, which solves the MHD equations in a general-relativistic context. These equations are defined as:
\begin{align}
        &\nabla_\mu (\rho u^\mu) = 0, \\
        &\nabla_\mu T^{\mu\nu} = 0, \\
        &\nabla_\mu \tensor[^\star]{F}{^{\mu\nu}} = 0,
\end{align}
where $\nabla_\mu$ denotes the covariant derivative, $\rho$ the rest-mass density, $u^\mu$ the fluid four-velocity, $T^{\mu\nu}$ the energy-momentum tensor (containing both ideal fluid and electromagnetic components), and $\tensor[^\star]{F}{^{\mu\nu}}$ the (Hodge) dual of the Faraday tensor. We assume an ideal MHD description, which implies that the electric field is acquired via the ``frozen-in'' condition (i.e., $\bm{E} = - \bm{v} \times \bm{B}$).

Our 3D simulations investigate three black hole spins, $a_\ast \in \{-0.9375, 0.0, 0.9375\}$, which will be referred to as \MM, \MZ, and \MP.
The simulations were performed using Modified Kerr-Schild coordinates \citep[MKS,][]{mckinney12}, which are logarithmically spaced in the radial direction and concentrates resolution in the equatorial plane (with $h=0.25$ and $R=0$, see Eqs. 49 and 50 in \citealt{porth17}). 
Nevertheless, we will list all following quantities in Kerr-Schild (KS) coordinates ($t$, $r$, $\theta$, $\phi$).

One of {\tt BHAC}'s main features is the inherent adaptive-mesh-refinement (AMR) construction of the underlying grid. Thus, dynamical refine- and derefinement is possible based on user-defined criteria. This allows for a more efficient resolution allocation and, therefore, increases computational performance. While the resolution is typically concentrated in the disk region to ensure a well-resolved magneto-rotational instability (MRI), our simulations have an additional static high refinement block on the jet-disk boundary.
As is shown explicitly in Appendix \ref{app:grid}, these blocks are at the highest AMR level corresponding to the resolution level that is listed in Table \ref{tab:GRMHDtable}, alongside other simulation parameters.
Due to the logarithmic radial spacing of the grid, this is still at a lower effective resolution level than close to be BH.
Another feature of our simulation suite is the high output cadence of 1 \rgc, which is a factor of 5-10 larger than comparable works.
This makes it well-suited to conduct variability and "slow light" studies (see both Sect. \ref{meth:slow_light} and \ref{res:slow_light}).

The simulations are initialized from a hydrostatic equilibrium solution \citep[FM;][]{fishbone76} with the main user-defined parameters and naming conventions being introduced in Table \ref{tab:GRMHDtable}.
The disk is seeded with a (perturbing) poloidal magnetic field loop that is initialized via the following vector potential \citep[see, e.g.,][]{wong21};
\begin{equation}
  A_\phi \propto \max\left( \frac{\rho}{\rho_\mathrm{max}} \left(\frac{r}{r_{\rm in}} \sin \theta \right)^3 e^{-r/400} - 0.2, \: 0 \right) . \label{eq:aphi}
\end{equation}
Here, $\rho$ and $r_\mathrm{in}$ denote the density and inner radius of the torus at initialization.
The resulting, evolved magnetic field brings about the MAD state which is characterized by magnetic flux eruption that locally and temporarily halt the accretion flow onto the black hole as demonstrated in Fig. \ref{fig:3DGRMHD}.
These flux eruption come about when the black hole saturates in horizon-penetrating magnetic flux after which the flux eruption generates a vertical magnetic field that creates a barrier and halts the accretion flow.
The resulting under-dense, over-magnetized regions are dubbed flux tubes \citep{mckinney12,dexter20,porth21}, as outlined more explicitly in Sect. \ref{int:introduction}.

\begin{table}
    \centering
    \caption{Overview of GRMHD simulations.}
    \begin{tabular}{ *5c } \toprule
        \emph{Model} & \emph{Spin} & \emph{Eff. Resolution} & $r_\mathrm{in}$ & $r_{\rho_\mathrm{max}}$  \\  
        \emph{Name} & $a_\ast$ & $N_r$, $N_\theta$, $N_\phi$ & & \\ \midrule
        {\tt MADP} & 0.9375   & 768,384,384 & 20 & 41 \\  
        {\tt MAD0} & 0.0      & 768,384,384 & 20 & 41 \\  
        {\tt MADM} & -0.9375  & 768,384,384 & 20 & 42 \\  
     \hline
    \end{tabular}
    \tablefoot{Additional simulation parameters are the adiabatic index $\hat{\gamma} = 13/9$, density floor $\rho_\mathrm{min}=10^{-4}$, magnetization ceiling $\sigma_\mathrm{max}=10^{3}$ for a simulation domain of $r \in [1.185, 3333.33]$, $\theta \in [0, \pi]$, $\phi \in [0, 2\pi]$ and a block size of $N_r \times N_\theta \times N_\phi = 8 \times 8 \times 8$ with up to four AMR levels. Two parameters that determine the size of the initialized disk are the inner radius ($r_\mathrm{in}$) and the location of the density maximum ($r_{\rho_\mathrm{max}}$).
    }
    \label{tab:GRMHDtable}
\end{table} 

\subsection{Radiative transfer} \label{meth:grrt}
To obtain synthetic images, light curves, and spectra, we post-processed our GRMHD simulations with the general-relativistic radiative transfer code {\tt RAPTOR} \citep{bronzwaer18,bronzwaer20}, which computes null-geodesics along which the radiative transfer equations are solved assuming thermal synchrotron emission (according to \citealt{leung11}).
{\tt RAPTOR}'s radiative transfer prescription takes into account both the synchrotron emissivity and absorptivity and is shown to be a robust numerical solver \citep{gold20,prather23}.
It is currently the only code that supports a direct readout of the native non-uniform (AMR) data structure as generated by {\tt BHAC} \citep{davelaar19}.
After the null-geodesics are calculated, we infer the electron temperature via a coupling relation as we are only evolving a (proton) one-fluid.
We utilize the so-called R$\beta$ description \citep{moscibrodzka16} that allows for a parametrization of the ion-to-electron temperature ratio ($T_\mathrm{ratio}$) as a function of the local plasma-$\beta$ (ratio of gas to magnetic pressure) from which we then acquire the dimensionless electron temperature ($\Theta_e$) according to:
\begin{align}
        &T_\text{ratio} = \frac{T_p}{T_e} = R_\text{low} \frac{1}{1 + \beta^2} + R_\text{high} \frac{\beta^2}{1 + \beta^2}, \\ 
        &\Theta_e = \frac{U \, (\hat{\gamma} - 1) \, m_p / m_e}{\rho \, (T_\text{ratio} + 1)}.
\end{align}
Hitherto undefined quantities are the plasma-$\beta = 2p/B^2$, internal energy density ($U$), proton mass ($m_p$), and electron mass ($m_e$).
In this work, we exclusively image our models with $R_\mathrm{high}=100$ and $R_\mathrm{low}=1,$\, which tends to produce more extended (i.e., jet-like) emission structures.
Generally speaking, a high $R_\mathrm{high}$ produces relatively hot jet-sheath electrons and relatively cool disk electrons, while for low $R_\mathrm{high}$, the electron temperature is more uniform.
We look at the accretion system under three inclinations angles $(i = 10^\circ, 30^\circ,  50^\circ$).
Our choices in $R_\mathrm{high}$, as well as the inclinations, are consistent with promising regions of the parameter-space as explored by \citet{eht22sgrav}. 
We evaluate the light curves and images at frequencies of $\nu \in \{19,21,23,25,30,32,34,36,41,43,45,47\}$ GHz \citepalias[following][]{brinkerink21}.
To appropriately cover the large-scale emission features present at these radio-wavelengths, we utilized a spherical polar camera grid, where the pixels are spaced logarithmically in the radial direction \citep[outlined in][]{davelaar18} with a resolution of $256 \times 256$ pixels, which has been shown to converge with results at higher resolutions \citep{davelaar18}, and a radial size of 500 \rg.
As regions in the inner jet-funnel tend to be dominated by emission from floor violations, it is necessary to exclude these from the radiative transfer procedure.
We employed both a geometrical and a standard (cold) magnetization criterion ($\sigma = B^2/\rho > 1.0$), for which we set the synchrotron emissivity to zero.
For the geometrical cut, we assumed a conical ($7.5^\circ$) cutout up to a radius $r=50$ \rg; afterward, we utilized a staged cylindrical (with cylindrical radius $r_c$) cutout (i.e., [i] $r_c < 10$ for $r \geqslant 50$ \rg, [ii] $r_c < 25$ \rg for $r \geqslant 300$ \rg).

Lastly, we introduced the relevant scaling relations.
Overall, GRMHD simulations are naturally scale-invariant and it is thus possible to scale it to any (RIAF) system. 
To scale our simulations, we need to define a black hole mass ($M_\mathrm{BH}$), which sets the time and length units, along with a user-defined parameter called the mass unit ($\mathcal{M}$) that sets the overall energy budget of the accretion flow. The total produced emission is also effectively set with $\mathcal{M}$ and this parameter is therefore varied until the desired integrated emission level is achieved.
We will predominantly be working with a theoretically advantageous unit-set that is based on gravitational radius $r_{\rm g} = G M / c^2$ and its time-unit $t_g = r_g/c$, which both reduce simply to $M$ in geometrized units $G=c=1$.
The black hole mass is set to $M_\mathrm{BH}=4.1\times10^6 \, M_\sun$ in combination with a distance to the central black hole of $D = 8.125$ kpc \citep{gravity_s2_18}. 
Expressed in cgs units, we find that $t_g \approx 20.2$ s and the angular size corresponding to a single $r_g$ is $\theta_{\rm g} = 5 \, \mu as$ \citep[see also, e.g.,][]{eht22sgrai}.

\subsection{Local cross-correlation function and goodness-of-fit estimation} \label{meth:lccf}

To compute the time lags from our synthetic light curves, we used the local cross-correlation function \citep{welsh99,max-moerbeck14}.
The LCCF enables cross-correlation between unevenly sampled timeseries and will be used exclusively to estimate the time lags in this work.
Even though the simulated light curves are evenly sampled by construction, we still utilized the LCCF as this is consistent with the methodology outlined in \citetalias{brinkerink15,brinkerink21}.
We cross-correlated all the light curves with a reference of $\nu_1$ = 19 GHz light curve, which implies that the auto-correlation will serve as the zeroth time-lag point. 
Overall, the choice of reference frequency does not significantly alter the found time lag relations, but some minor variations are perceivable, as is outlined in detail in Appendix \ref{app:different_reference_freq}.

To determine the similarity between the simulated ($E_{\mathrm{sim}}$) and observed time lag ($O_{\mathrm{obs}}$), we evaluated the $\chi^2$ statistic for the six listed time lag values (excluding 100 GHz) in \citetalias{brinkerink15} to see which light curve sections are most similar to the observational equivalent.
However, as the acquired time lags can be arbitrarily shifted ($\xi$) in time, we minimized the following function to assess goodness of fit: 
\begin{equation}
  \chi^2 = \sum_i^{N} \left[ \frac{O_{\mathrm{obs},i} - E_{\mathrm{sim},i} - \xi}{\sigma_{\mathrm{obs},i}} \right]^2 ,
\end{equation}
where $\sigma_\mathrm{obs}$ denotes the observational standard deviation and the sum is over the number of frequencies, $N=12$.
In addition to the goodness-of-fit estimation with the observations directly, we also calculated it with respect to the linear fit from \citetalias{brinkerink15} (with a slope of $A = 41 \pm 14$ cm/min) to gauge the similarity to the preferred linear relation, as expected from theory \citep{falcke09}.
We note that we calculate the $\chi^2$ estimates in a sliding window fashion, which is determined by the starting time and window length.
A tabulated overview of the best-fitting windows is listed in Appendix \ref{app:bestwindows}.
These windows are also displayed with horizontal (grey and black) bars in Figs. \ref{fig:MPlightcurve+CC}, \ref{fig:M0lightcurve+CC}, and \ref{fig:MMlightcurve+CC}.

\subsection{Favored linear fit} \label{meth:linfit}

With the analysis explained here, we have a two-fold objective, namely to assess (i) how linear the acquired time lag relation is and (ii) how well it corresponds to the linear fit acquired in \citetalias{brinkerink15} (with a slope of $A = 41 \pm 14$ cm/min). 
As we know from isotropic outflow arguments, a linear relation is expected in the case of a relatively simplistic conical jet structure \citep{falcke09} and this serves as an interesting test to find the deviations from this picture.
Thus, more specifically, we will fit a linear relation ($y = Ax + B$) to all time lags acquired from the simulations and assess the root-mean-square error (RMSE) with a linear fit to this relation to evaluate goodness-of-fit according to:
\begin{equation}
    \mathrm{RMSE} = \sqrt{\frac{1}{N} \sum_{i=1}^{N} \left(T_i - T_{\mathrm{f},i}\right)^2}.
\end{equation}
Here, $T$ is the simulation time lag and $T_\mathrm{f}$ is the time lag relation prescribed by the fitted linear relation for the ($N=$)12 frequencies evaluated in this work. Typically, we would find $\mathrm{RMSE} \lesssim 5$ for time-lag windows that show a reasonably linear relation. 

\subsection{Slow light implementation} \label{meth:slow_light}
The fast- versus slow-light paradigm has been quite  widely investigated for mm-wavelength emission originating close to the BH \citep[see, e.g.,][]{dexter10,bronzwaer18,moscibrodzka21}.
For ray-tracing GRMHD simulations, the standard in the field is to use a ``fast light'' prescription, which implies that the plasma description is not evolved; whereas the radiative transfer equations are evaluated along the null-geodesics.
This effectively renders the speed of light to be infinite.
For a more realistic description, denoted as ``slow light,'' we needs to evolve the plasma and light simultaneously.
Overall, we can identify two scenarios where a slow-light prescription could make a significant difference.
First, this effect can be important when a strong space-time curvature present, which predominantly indicates emission originating from within the inner $5-10$ \rg.
Even though matter moves fast ($v \cong 0.5c$) for this scenario, relatively few photons escape from the innermost accretion regions and we typically recover only minor (or even negligable) time lags when only near-horizon emission features are evaluated, which is partly explained by the confined region in which they are created. 
Second, when matter is moving (relatively) fast over long distances, which we associate with emission from the jet(-sheath).
Contrary to all previously mentioned studies at $230$ GHz (1.3 mm), our scenario falls within the second category, which is the more complex case as we will outline in the following paragraph.

The complexity of this latter case lies predominantly with the large-scale and diffuse emission structure that presents itself, for example, in the 19 GHz images in this work.
So, when the emission structure is compact (e.g., at 230 GHz), then we would only have to read in a relatively low number of GRMHD snapshots ($\mathcal{O}(50)$) to get a consistent picture; typically, the number of required snapshots is comparable to the size of the mission structure.
To determine the required number of snapshots, we performed a convergence study that indicates that slow-light images at 19 GHz converge for $\sim$600 read-in GRMHD snapshots, where we used a 1 \rgc time cadence.
This convergence test was conducted by cross-comparing images with an increasing amount of read-in snapshots, where the maximal case was calculated with 1200 read-in GRMHD snapshots.
Increasing the snapshot number higher than $\sim$600 results in minor $\mathcal{O}(0.5\%)$ differences in the images and the light curves. 
For the remainder of the slow light analysis, we therefore used 600 snapshots.
Naturally, the convergence number is (partially) determined by when one reaches the limiting optical depth with $\tau_\nu > 1,$ after which the medium becomes (too) optically thick and the integrated flux density no longer increases.
This is the case for all radiative transfer studies, however, it is only for slow light that this limiting surface is dependent on the plasma conditions over a given time period. 

In practice, we employed a hybrid slow and fast light approach, where we calibrated the slow-light integration region along the geodesic so that (almost) all emission is described in the slow light window.
As mentioned, for this slow light implementation, we read-in 600 GRMHD files (${\sim}350$ GB in RAM) that will give the plasma description at different times along the geodesic and it is therefore a memory-intensive procedure.
The implementation relies on two dependent parameters, $r_\mathrm{slow}$ and $N_\mathrm{snap}$, while the latter denotes how many GRMHD snapshots are read in, the former determines from which distance (to origin) the slow light description is assumed.
As the geodesics are integrated backwards, the GRMHD description starts to move back in time until it has reached $T_\mathrm{END} = T_\mathrm{START} - N_\mathrm{snap}$.
All integration before $T_\mathrm{START}$ and after $T_\mathrm{END}$ assumes fast light.
For this approach, we have to calibrate for $N_\mathrm{snap}$ and $r_\mathrm{slow}$ to find the number after which differences in emission structure are at an acceptably minimal level.
In this work, we chose to include a study of the most prominent fast light time window, as we are investigating a scenario where a slow light description could have a significant impact (Sect. \ref{res:slow_light}).

\section{Results} \label{res:results}
In the following sections, we discuss the fitted spectra (Sect. \ref{res:SED}), the qualitative (Sect. \ref{res:qualitative}) and quantitative (Sect. \ref{res:quantitative_time_lags}) aspects of the time lag analysis, and the implications of adopting a slow light radiative transfer approach (Sect. \ref{res:slow_light}).

\begin{figure*}
    \centering
    \includegraphics[width=\textwidth]{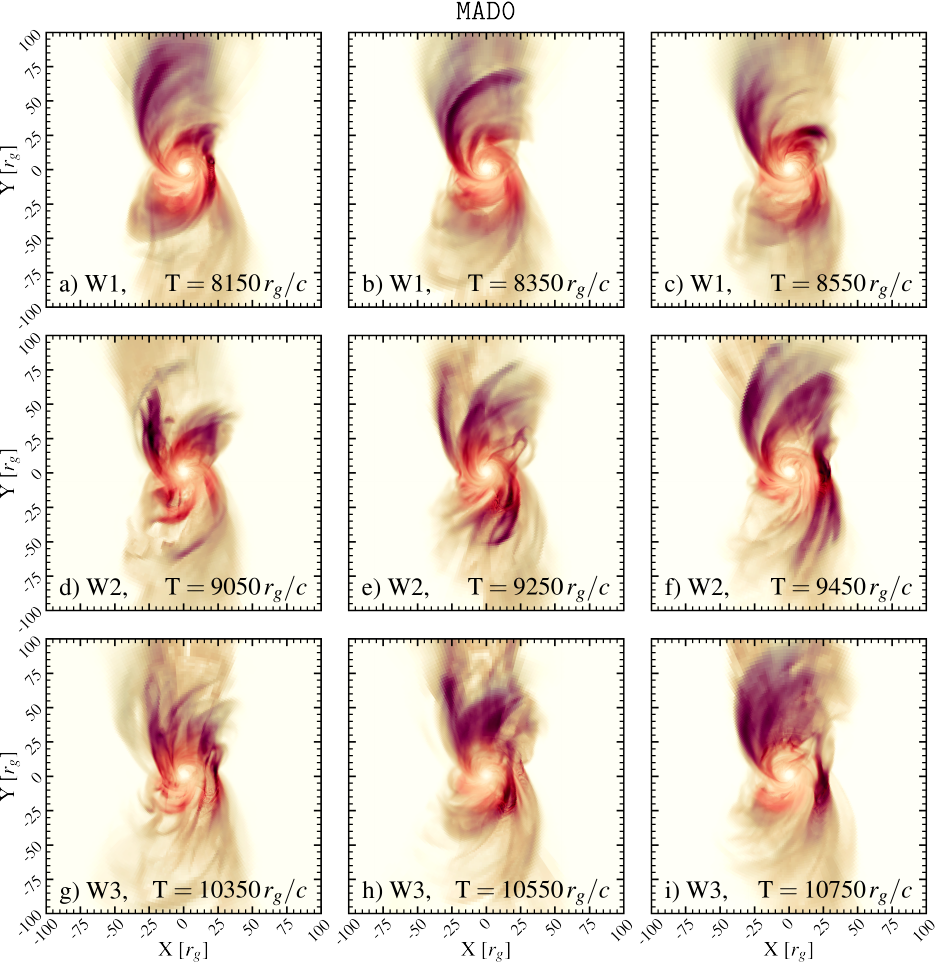}
    \caption{Ray-traced images for inclination $i=30^\circ$. The images consist of three combined colormaps corresponding to the thermal synchrotron emission at $\nu = 19, \, 32, \, \mathrm{and} \: 47$ GHz in shades of green-blue, red-purple, and orange-red, respectively. When emission from all frequencies is present, one creates a black-purple hue in the composite image. All base image maps describe the flux density ($F_\nu$) between $0.0$ and $0.025$ Jy per pixel. The base images are combined with equal weight to acquire the composite images. More details on the procedure as well as animations for all $i=30^\circ$ cases are provided in Appendix \ref{app:animations}. 
    }
    \label{fig:grrtmultichrome}
\end{figure*}

\subsection{Spectral energy distributions, accretion rates, \newline and jet power estimates} \label{res:SED}

Figure \ref{fig:SED} displays the spectral energy distributions (SEDs) for all our ray-traced models.
The spectra were fitted with a binary search algorithm \citep[also utilized and outlined in][]{bronzwaer21}, which launches the ray-tracing code in an iterative manner and fits the time-averaged (over $T \in [6300,15300] \, r_\mathrm{g}/c$) 230 GHz flux density $F_\nu$ to $2.5 \pm 0.03$ Jy.
With this relatively simple procedure and an exclusively thermal electron population, we obtained good fits to the quiescent \sgr spectrum at inclinations of $i=10^\circ$, $30^\circ$, and $50^\circ$.
Especially in the cases of $i=30^\circ$ and $i=50^\circ$, we obtained SEDs that correspond well to the binned observational data that are gathered from \citet{falcke98,schodel11,bower15,bower19,witzel18,vonfellenberg18,gravity20}.
At $i=10^\circ$, we tend to either under- or overproduce the longer wavelength radio emission (i.e., $\nu\lesssim100$ GHz).

For our fits, we find a corresponding accretion rate via $\dot{M}_\mathrm{cgs} = \mathcal{M} \cdot \langle \dot{M}_\mathrm{sim}$\footnote{Where $\dot{M}_\mathrm{sim} = \int_0^{2\pi} \int_0^{\pi} \rho u^r \sqrt{-g} \mathrm{d}\theta \mathrm{d}\phi$ is the horizon-penetrating mass flux \cite[cf.][]{porth19}.}$\rangle / t_\mathrm{g}$
, where the simulation accretion rate ($\dot{M}_\mathrm{sim}$) is the averaged over the evaluated time-domain (i.e., $T \in [6300,15300] \, r_\mathrm{g}/c$).
For inclination $i=30^\circ$, we find accretion rates of $\dot{M}_\mathrm{cgs} = 5.69 \times 10^{-9}, 1.52 \times 10^{-8}, 1.62 \times 10^{-8} \, M_\sun / \mathrm{yr}$ for \MP, \MZ, and \MM, respectively.
As $\mathcal{M}$ is calculated for every inclination (and simulation) separately, the accretion rate varies by several factors, but it remains low, as is outlined in Appendix \ref{app:accretion_rate}.
Here, we list the $i=30^\circ$ accretion rates explicitly, as they correspond to the favored models in this work.

As some of the classical time lag interpretation finds its origin in jetted outflow models, it could be informative to gauge the jet power of our GRMHD models.
Following the methodology outlined in (detail in Appendix A of) \citet{eht5}, we can estimate the jet power ($P_\mathrm{jet}= \langle P_\mathrm{jet,sim}$\footnote{Where $P_\mathrm{jet,sim} = \int_{(\beta\gamma)^2>1} \mathrm{d}\theta \int_0^{2\pi} \mathrm{d}\phi \sqrt{-g} (-\tensor{T}{^r}{_t} - \rho u^r)$, which is outlined in detail in \citet{eht5}.}$\rangle \cdot \mathcal{M} c^2$
)
of our GRMHD simulations (averaged over the evaluated time-domain, $T$), which were calculated to be $P_\mathrm{jet} = {\sim}10^{39}, {\sim}10^{37}, {\sim}10^{39}$ erg/s at a radius of $r=100$ \rg for \MP, \MZ, and \MM, respectively.
We found only minor variations in $P_\mathrm{jet}$ based on inclination ($i$).
These estimates are consistent with what was presented previously for Sgr A$^\ast$ \citep{eht22sgrav}.
As no robust jet feature has been observed for Sgr A$^\ast$ (yet), it is hard to gauge if these values are realistic \citep[see discussion in][and references therein]{eht22sgrav}.
We do note that our interpretation of time lags does not hinge on the presence of a clear or strong jet signature; rather, it focuses on the complex magnetically driven dynamical interactions, as we outline below.

\subsection{Qualitative analysis of time lags} \label{res:qualitative}

Figure \ref{fig:grrtmultichrome} displays composite ray-traced images of thermal synchrotron emission at $19$ (blue-green), $32$ (pink-purple), and $47$ GHz (red-orange) for selected windows of the \MZ case.
This allows for a more intuitive interpretation of how emission features move through frequency space, where dark regions indicate coincident emission at all frequencies.
As time lags generally move from high to low frequencies, it will start out as a red hue and gradually move to a dark purple hue indicating coincidence of multiple (lower) frequencies as is illustrated in Appendix \ref{app:animations}.
Window 2 (W2, in panels $d$,$e$,$f$) gives a prime example of this behavior, especially if one looks at the flux tube that starts out with a red color and gradually becomes darker as it cools (which is illustrated explicitly in Appendix \ref{app:grid}). 

Figures \ref{fig:MPlightcurve+CC}, \ref{fig:M0lightcurve+CC}, and \ref{fig:MMlightcurve+CC} display the corresponding light curves and LCCF coefficients of three sections of the light curves (W1, W2, and W3) for the \MP, \MZ, and \MM cases, respectively. 
These windows are arbitrarily chosen to represent a variety of time lag relations as found for our simulations (detailed statistics are calculated in Sect.~\ref{res:quantitative_time_lags}).
The classical interpretation of the time lags rests on the frequency-dependent synchrotron emission signature from a (mildly) relativistic, conical jet \citep[see, e.g.,][]{blandford79,falcke93a,falckebiermann99iii,falcke09}, or even an expanding blob of plasma \citep{vanderlaan66}.
Even though the jet picture is broadly correct, it is not sufficient to explain the best-fitting time-lag windows from our simulations.
As these are relatively short observing windows (of up to $400 \, r_g/c \approx 2.25 \, \mathrm{hrs}$), the variable emission component is arguably more important than the relatively static (jet) emission structure, which is confirmed by our findings. 
More specifically, we note that a number of windows (in Figs.~\ref{fig:MPlightcurve+CC}, \ref{fig:M0lightcurve+CC}, and \ref{fig:MMlightcurve+CC}) contain good time-lag fits and follow closely after a flux eruption which is denoted by the drop in $\Phi_\mathrm{B}$ (magenta line), as will be explained in more detail in next paragraph(s). 

We postulate that the main driver behind the variable component in our MAD simulations is the formation of flux tubes after a magnetic flux eruption.
During these flux eruptions, a low-density, high-magnetization region is created after the BH saturates in horizon-penetrating magnetic flux \citep[$\Phi_\mathrm{B}$;][]{tchekhovskoy11}.
When this occurs, a part of the magnetic flux $\Phi_\mathrm{B}$\footnote{Where $\Phi_B = \tfrac{1}{2} \int_{0}^{2\pi} \int_{0}^{\pi} |\tensor[^\star]{F}{^{rt}}| \sqrt{-g} \, \mathrm{d}\theta \, \mathrm{d}\phi$ is the horizon-penetrating magnetic flux \citep[cf.][]{porth19}. The limiting value of the MAD-parameter $\phi = \Phi_B / \sqrt{\dot{M}} \approx 15$ in our unit-set.} is dissipated by means of a magnetic reconnection event that generates a strong vertical magnetic field component that partly halts and pushes back the accretion flow effectively creating the flux tube.
After this event, the horizon-penetrating magnetic flux $\Phi_B$ (magenta line) drops and seems to strongly correspond with the overall variation of the light curves, especially for the \MP case and, to a lesser extent, for the \MM case.
The flux tube is aligned with the jet and orbits the BH at sub-Keplerian velocities before expanding and eventually dissipating into the disk.
This can span several orbital periods ($\sim 1500$ \rgc for, e.g., W2 in Fig. \ref{fig:M0lightcurve+CC}). 
At its birth, it is predominantly visible in the higher frequency emission (red) before it expands and cools so it will start emitting at lower frequencies also (creating the purple-black hue).

Interestingly, as it matures, the tail-end of the flux tube (furthest away from the BH) produces the strongest emission feature as the flux tube is pushing against the accretion flow which compresses (creating an over-density) and heats the plasma as is clearly shown in Appendix \ref{app:grid} \citep[also commented on in][]{mahdi23}. 
This emission feature resembles a ``hot spot'' \citep{broderick06,vos22}, especially when the flux tube is receding with respect to the observer as one then directly looks at (the back of) the compression region.
We note that this picture deviates in interpretation from the standard hot spot model \citep[cf.][]{vos22}.
There, the point of maximal emission typically occurs when the spot approaches the observer and is maximally (Doppler) beamed.
For $i=30^\circ/50^\circ$, however, we can clearly see a vertical tubular emission structure, which is quite different from the typical spherical hot spot picture.
Another channel by which variable emission features are introduced is the non-homogeneous nature of the jet sheath (or wall), which produces the majority of the emission overall.
Over-dense outflows will occur sporadically and are sheared and subsequently ``smeared out'' over the jet sheath cone.
Nevertheless, we find that the jet-sheath-related bursts in emission are often closely preceded by a flux eruption event; so, even though they originate is different parts of the simulation domain, they are triggered by the same physical mechanism.
While it is typically hard to pinpoint the main emission contributor to well-corresponding time lags, it is striking that flux tubes are prominently featured in the best windows (as can be confirmed with the animations provided in Appendix \ref{app:animations} together with the horizontal markers in Figs. \ref{fig:MPlightcurve+CC}, \ref{fig:M0lightcurve+CC}, and \ref{fig:MMlightcurve+CC}).

\begin{figure*}
    \centering
    \includegraphics[width=\textwidth]{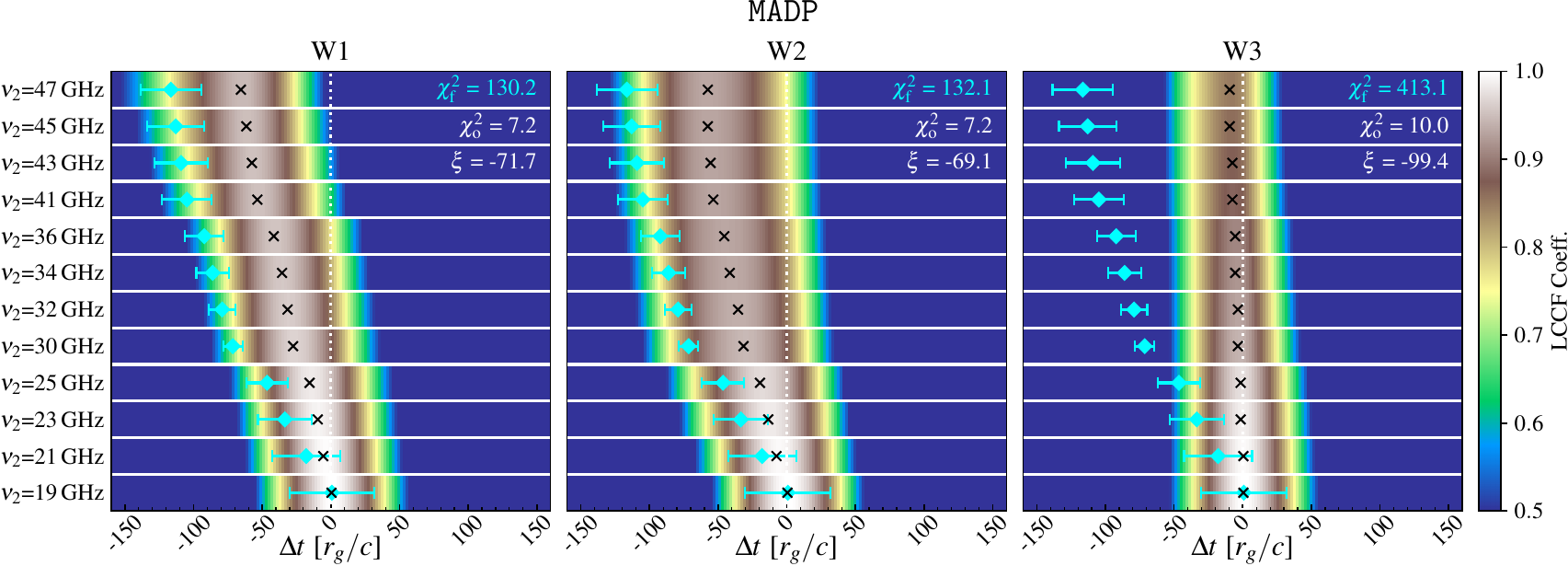}
    \includegraphics[width=\textwidth]{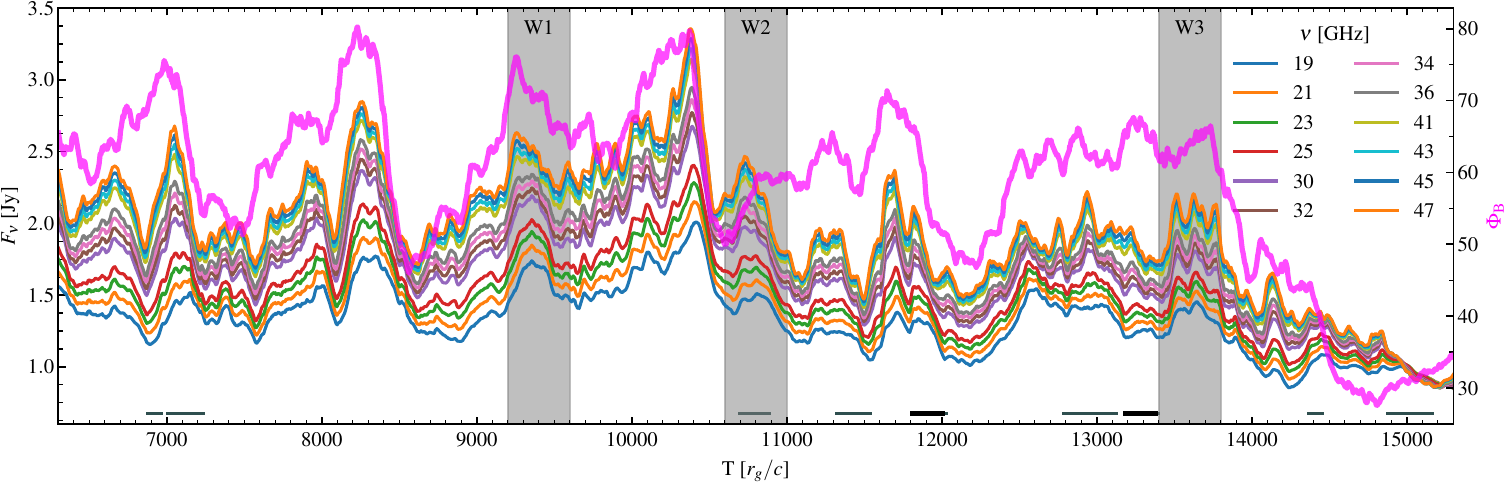}
    \caption{LCCF coefficients and light curves for the \MP case at $i=30^\circ$. For the LCCF calculations (\textit{top}), we chose one light curve ($\nu_1 = 19$ GHz) that we cross-correlated with all other light curves ($\nu_2$). The black crosses (``x'') denotes the maximum of the LCCF within a given window. The cyan diamants with errors denote the fit obtained in \citetalias{brinkerink15}. Both  light curves, ranging from $19$ to $47$ GHz, and horizon-penetrating magnetic flux $\Phi_B$ (right-hand axis) are shown in the bottom panel. The three windows W1, W2, and W3 are $400$ \rgc in length and correspond to the LCCF panels shown in the top. The grey and black horizontal bars denote the best-fitting time-lag windows (as prescribed by criteria (iii) and (iv), respectively, in Appendix \ref{app:bestwindows}). }
    \label{fig:MPlightcurve+CC}
\end{figure*}

\begin{figure*}
    \centering
    \includegraphics[width=\textwidth]{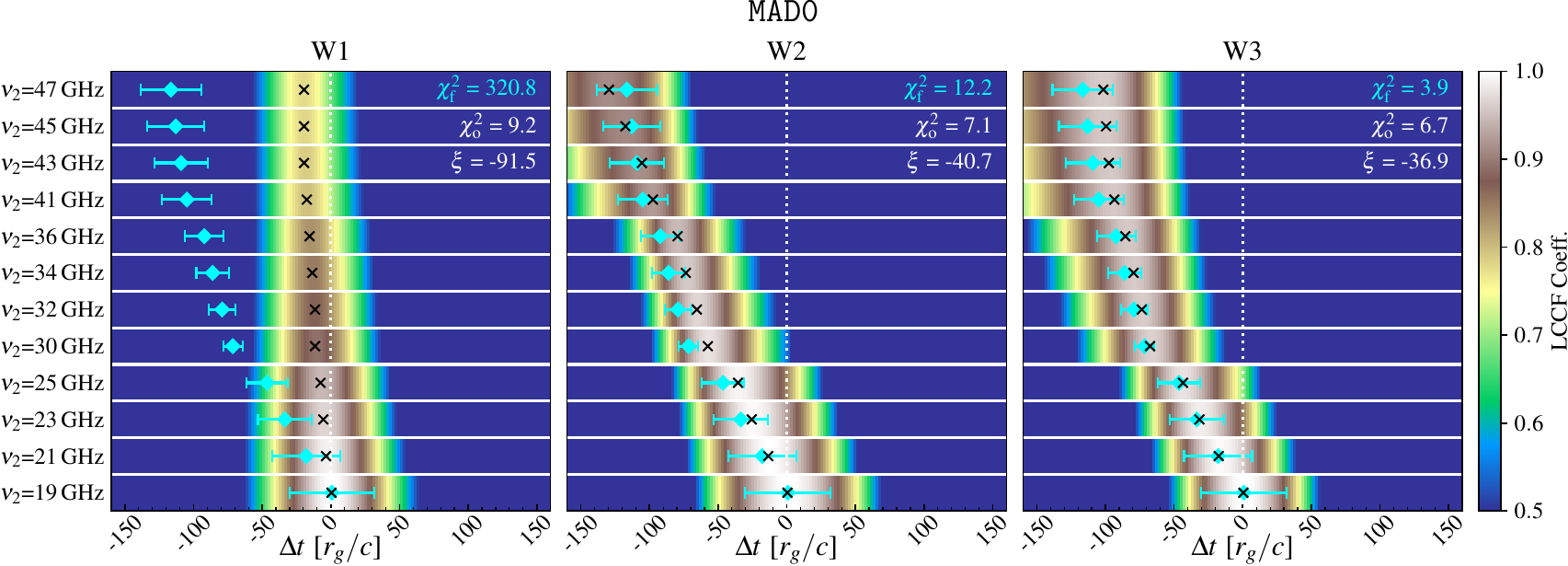}
    \includegraphics[width=\textwidth]{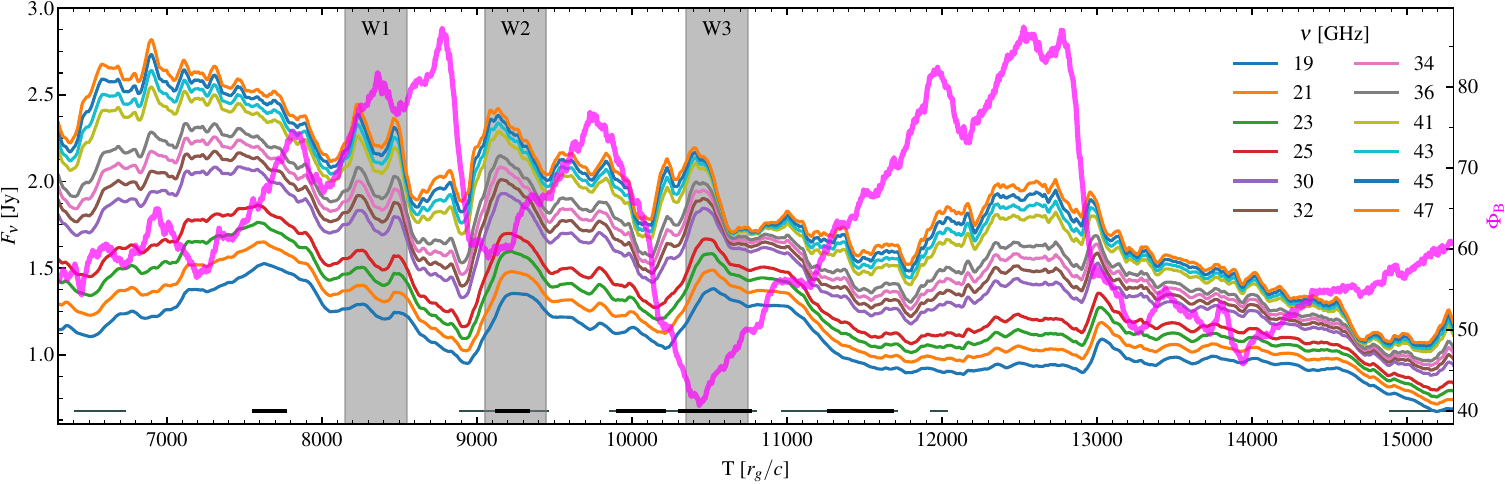}
    \caption{LCCF coefficients and light curves for the \MZ case at $i=30^\circ$. The rest of the description of Fig. \ref{fig:MPlightcurve+CC} is also applicable here. }
    \label{fig:M0lightcurve+CC}
\end{figure*}

\begin{figure*}
    \centering
    \includegraphics[width=\textwidth]{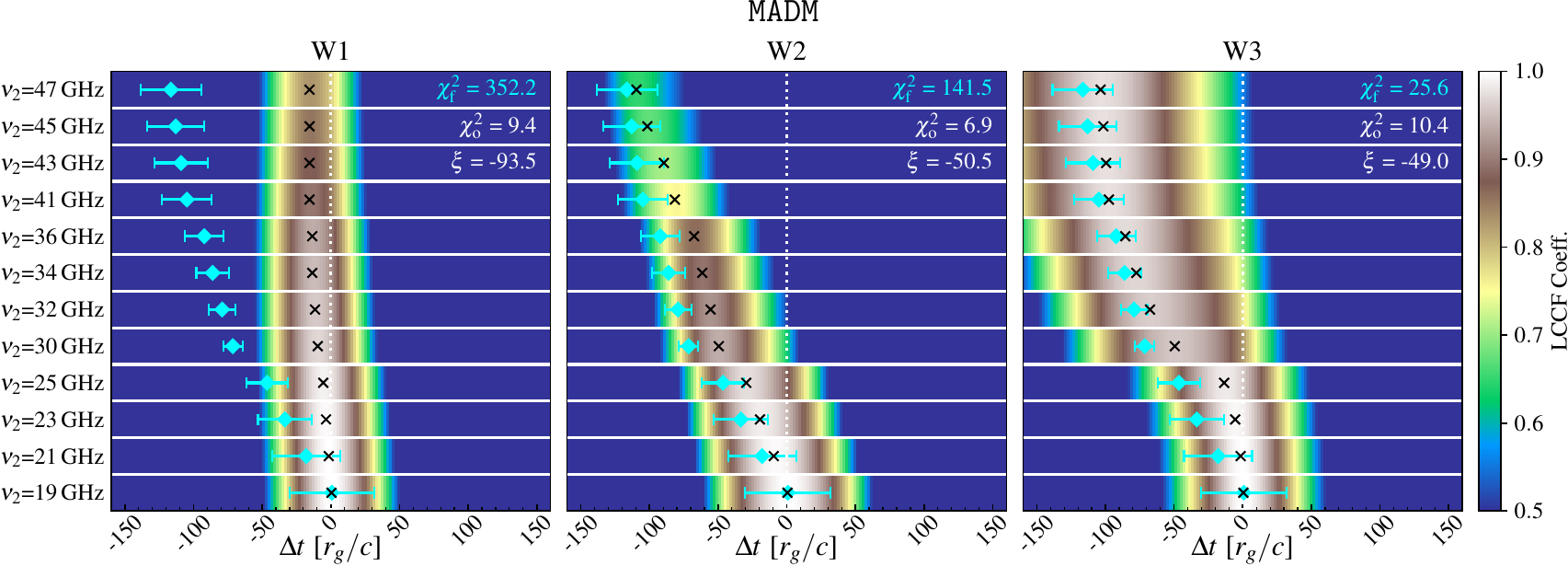}
    \includegraphics[width=\textwidth]{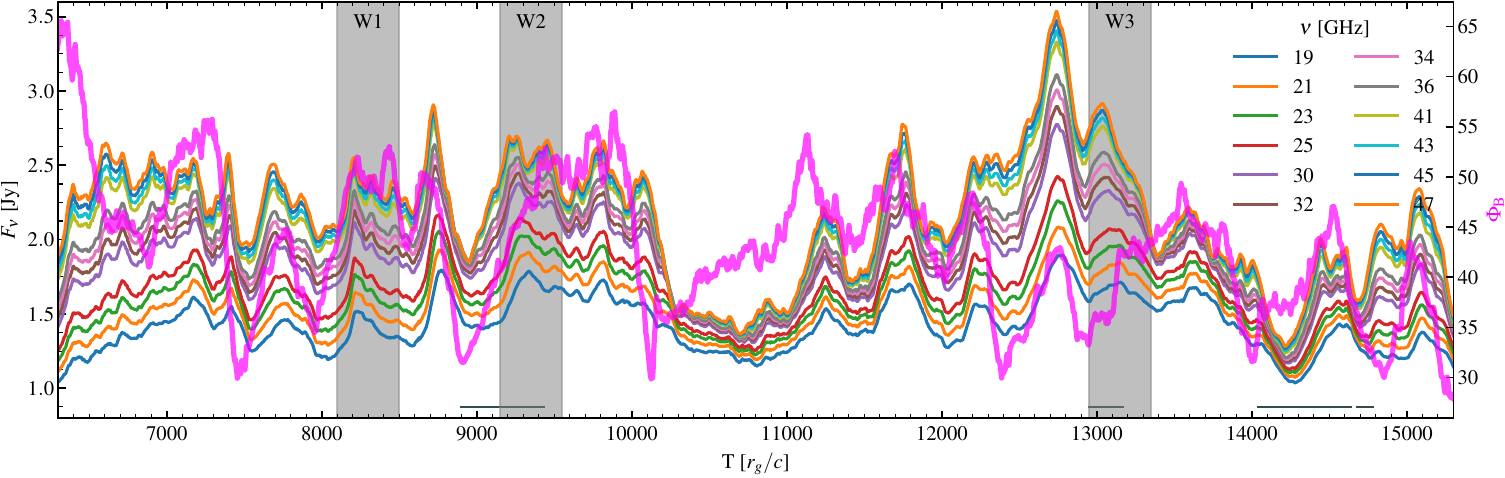}
    \caption{LCCF coefficients and light curves for the \MM case at $i=30^\circ$. The rest of the description of Fig. \ref{fig:MPlightcurve+CC} is also applicable here. }
    \label{fig:MMlightcurve+CC}
\end{figure*}

\subsection{Quantitative analysis of time lags} \label{res:quantitative_time_lags}

\begin{figure*}
    \centering
    \includegraphics[width=\textwidth]{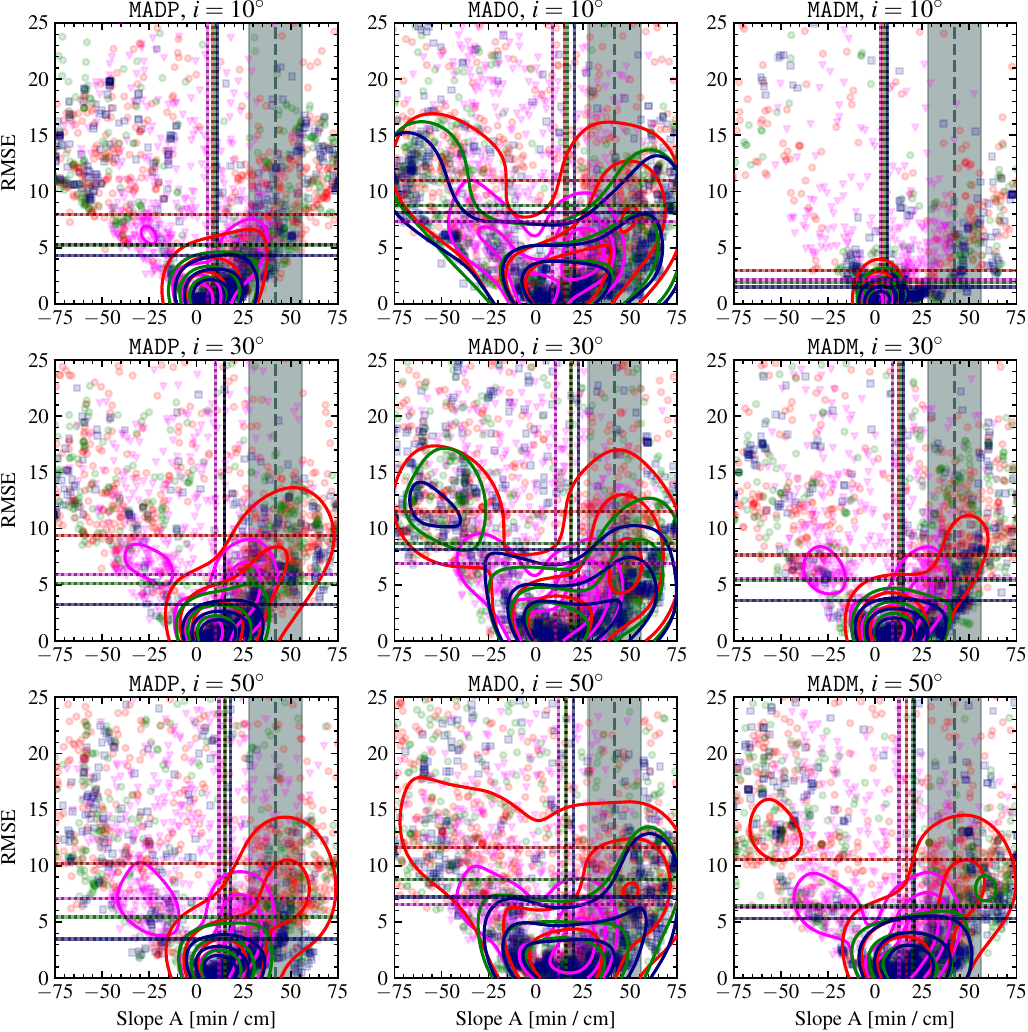}
    \caption{
    RMSE between the time lag acquired from the simulations and their linear fit. 
    The points are color-coded according to the time-lag window length; 100 (\emph{magenta}), 200 (\emph{red}), 300 (\emph{green}), and 400 (\emph{blue}) \rgc.
    The color-dashed lines denote the mean slopes of the best-fitting (RMSE < 5) time lags for the vertical lines, while the horizontal lines denote the mean RMSE of the entire population.
    The grey dashed line, with standard deviation area, corresponds to the linear fit to the VLA-only data from \citetalias{brinkerink15}, which is $A = 42 \pm 14$ cm/min. The contours denote the $25\%$, $50\%$, and $75\%$ levels of the normalized kernel density estimations for each of the window length results (in the corresponding colors).
    }
    \label{fig:slopeapp}
\end{figure*}

To quantify how well our simulations agree with the findings of \citetalias{brinkerink15,brinkerink21}, we calculated the time lags over sections of the synthetic light curves for time windows of 100, 200, 300, and 400 \rgc.
These sliding windows are shifted with steps of 10 \rgc until the entire light curve is covered.
The acquired time lags are then scored with two types of $\chi^2$ estimate.
Following the method outlined in Sect. \ref{meth:lccf}, we calculated $\chi^2_\mathrm{o}$ directly with the observations.
Then, we calculated $\chi^2_\mathrm{f}$ between the linear fits listed in \citetalias{brinkerink15} and our simulated ones to get an indication of whether we are able to recover this slope.
The overview in Table (\ref{tab:chi2}) is listed in Appendix \ref{app:bestwindows}.
We note that the 25 GHz measurement \citepalias[by][]{brinkerink15} deviates significantly from the otherwise linear trend in the data and is therefore not well represented in the linear fit.
We also investigate the preferred linear relations based solely on our simulated data.
This will be outlined later in this section and displayed in Fig. \ref{fig:slopeapp}.
Overall, $\chi^2_\mathrm{o}$ is significantly less constraining than $\chi^2_\mathrm{f}$ as a result of the large errors on the observational time lags.
Nevertheless, the combined constrains (both $\chi^2 < 7$) are restricting, but still identify a number of passing windows with a preference for the \MZ case, especially with $i=30^\circ/50^\circ$ and longer correlation windows.
The choice for $\chi^2 < 7$ is arbitrary and quite stringent, but it also clearly highlights windows where the simulation behavior is highly consistent with \citetalias{brinkerink15}, as we find a reduced $\chi^2_\nu \lesssim 1.2$ (for the six observational points in the case of $\chi^2_\mathrm{o}$ and 12 points for $\chi^2_\mathrm{f}$ as shown in, e.g., Fig. \ref{fig:MPlightcurve+CC}). 

The physical manifestation of the best-fitting time lags corresponds to a flux tube that starts directly in front of the main jet structure and continues its counter-clockwise (as seen from $i=0^\circ$) trajectory until it moves behind the jet structure.
Even behind the jet, the flux tubes still contributes to the observed flux.
Both these points are clearly demonstrated in the provided animations in Appendix \ref{app:animations}, especially when combined with an evaluation of the light curves displayed in Figs. \ref{fig:MPlightcurve+CC}, \ref{fig:M0lightcurve+CC}, and \ref{fig:MMlightcurve+CC}.
As clearly seen in W2 and W3 of Fig. \ref{fig:grrtmultichrome}, the flux eruption that creates the flux tube also give an increase in the high frequency (red) emission that will gradually move to lower frequencies in the jet sheath (dark purple).
The time lag displays quite an characteristic kink in the low frequency bands to accommodate for the outlier (at 25 GHz). 
The largest passing window is broadly denoted by W3 for the \MZ case, as is shown in both Figs. \ref{fig:grrtmultichrome} and \ref{fig:M0lightcurve+CC}.
For completeness, we note that the data listed in \citetalias{brinkerink21} is more complex in nature than what was used for \citetalias{brinkerink15}, but is still consistent with the linear relation that we investigate (i.e., $\chi^2_\mathrm{f}$ in Table \ref{tab:chi2}).

Figure \ref{fig:slopeapp} answers two main questions; how linear are the simulation time lag relations and which slope would best fit them?
While all distributions display considerable variance, we note that one finds quite a considerable number of points in the area corresponding to the fit of \citetalias{brinkerink15} in the grey shaded area (for $A = 41 \pm 14$~cm/min).
The mean slopes of the best-fitting time-lag distributions (denoted by the black-color dashed vertical lines) are significantly lower than the observational slope.
This indicates that the most linear time lags from the simulations favor a more shallow slope than was found for \citetalias{brinkerink15} or that the steeper (better fitting) time lags do not have a (very) linear profile.
Therefore, based on the results presented here, we find that the relatively steep slope is a rare occurrence according to our simulations.
A last point to consider is that a significant portion of observational time lags do not show a clear relation \citepalias{brinkerink21}. 
Interestingly, this behavior is also recovered for our idealized simulations as they do regularly not show a clear time lag relation (i.e., there are breaks or multiple competing features in the LCCF coeff.), as is indicated by the black dashed horizontal lines that denotes the mean RMSE of all points.
Overall, based on the kernel density contours shown in Fig. \ref{fig:slopeapp}, we find that most of the simulated time lags show a predominantly linear ($\mathrm{RMSE} < 5$) trend that is typically lower than $A = 41 \pm 14$~cm/min.
The linear trend that is preferred from simulations is $A \in [7,23]$~cm/min based on the $400$ window length calculations, with $A \approx 20$~cm/min for \MZ and lower values for \MP and \MM, which increase with inclination from $A \approx 10$~cm/min at $i=10^\circ$ to $A \approx 20$~cm/min at $i=50^\circ$. 
These listed values are denoted by the vertical lines in Fig. \ref{fig:slopeapp}.
 
\subsection{Slow light comparison} \label{res:slow_light}
After one has acquired the slow light curve, it is beneficial to find the optimal offset between both the fast and slow light curves.
This offset can then be applied to one of the time series to allow for intuitive comparison. 
However, finding the offset can be a somewhat non-trivial procedure as the timing between various features differs depending on the specific light description.
While there are several options available to find the offset (e.g., via cross-correlation of either the images or the light curves), we have have opted for the minimization of the Euclidean distance between segments of the slow and fast light curves (i.e., 1D equivalent of the down-hill-simplex method; \citealt{press02}).
Taking $\nu = 19$ GHz as the reference frequency, we find an offset in time of $\Delta t = 365$ \rgc, while for $\nu = 47$ GHz, we find $\Delta t = 402$ \rgc.
Nevertheless, we would like to note that for the outcome of the (LCCF) time lag calculations, the alignment shift is inconsequential as we are calculating the time lags between the various frequency light curves within their corresponding groups.

\begin{figure*}
    \centering
    \includegraphics[width=\textwidth]{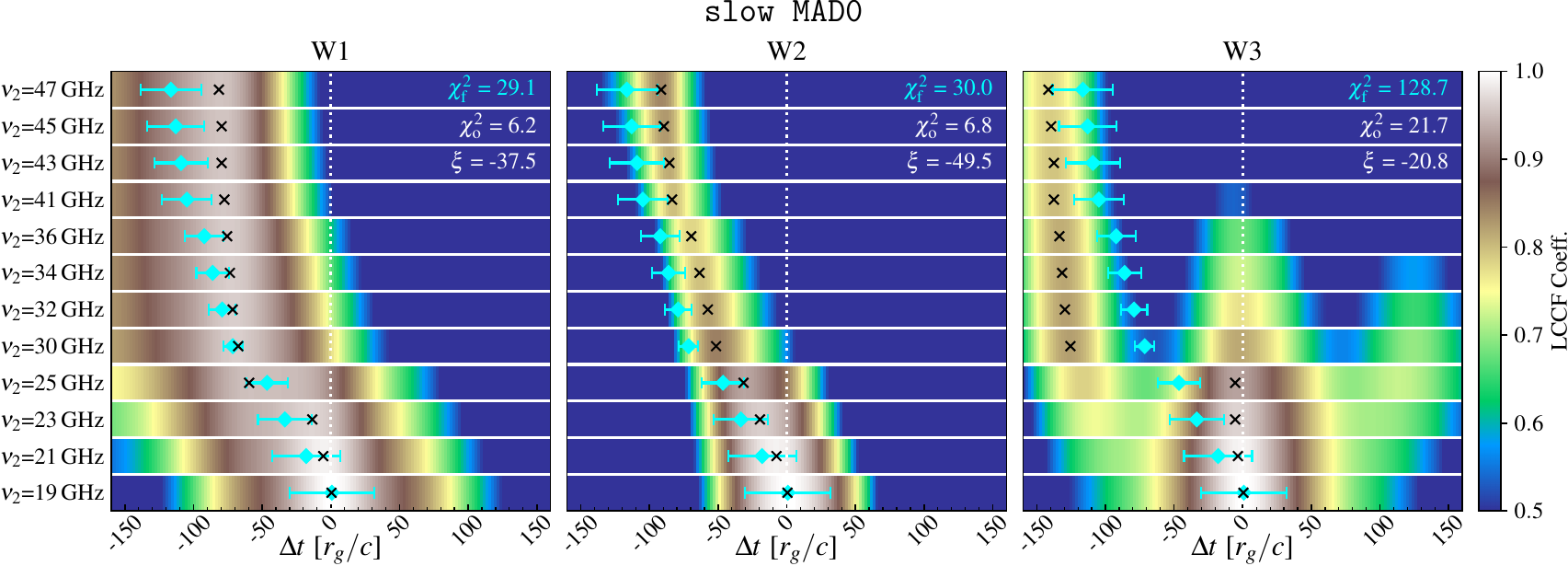}
    \includegraphics[width=\textwidth]{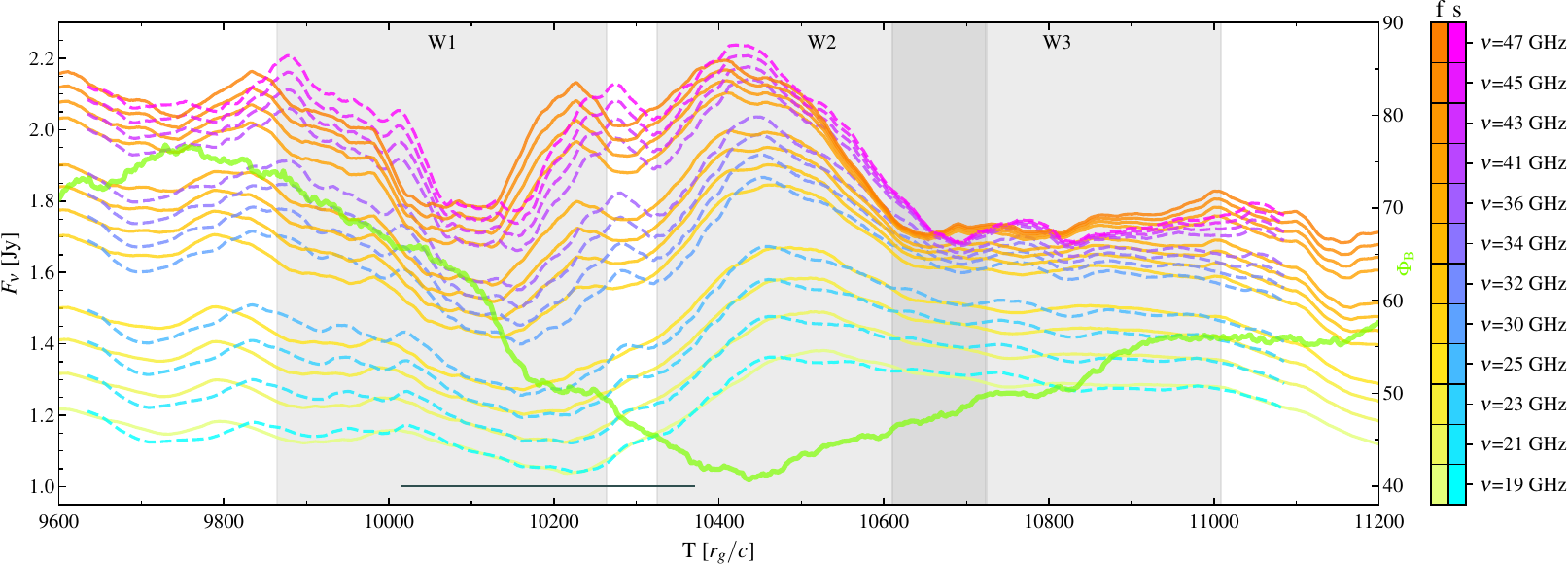}
    \caption{Fast and slow light curves for the \MZ case at $i=30^\circ$. The slow light window was chosen to correspond to the most promising time-lag window for the fast light curves, as indicated by Table \ref{tab:chi2}. The color-schemes corresponding to the various frequencies for the fast (\emph{solid}) and slow (\emph{dashed} lines) light curves are denoted by ``f'' and ``s'', respectively. 
    The three windows W1, W2, and W3 are 400 \rgc in length and correspond to the LCCF panels (calculated for the slow light curves) shown at the top, analogously to Fig. \ref{fig:M0lightcurve+CC} for fast light. The horizon-penetrating magnetic flux ($\Phi_\mathrm{B}$) is denoted in green (correspond to the right axis) and the horizontal (\textit{grey}; only for criterion (iii) as outlined in Appendix \ref{app:bestwindows}) bars denote the best-fitting time-lag regions.
    }
    \label{fig:slowfastlightapp}
\end{figure*}

Figure \ref{fig:slowfastlightapp} displays the fast and slow light curves corresponding to the promising window centered around $T = 10550$ \rgc for the \MZ, $i=30^\circ$ case.
What becomes clear is that the slow light description does have a non-negligible effect on resulting time lags. 
We only calculated the alignment shift for $\nu = 19$ GHz and shifted all slow light curves accordingly.
As we expected, timing between various light curve features has changed significantly and results in a relatively worse correspondence with the observational time lags (especially for $\chi^2_\mathrm{f}$, as listed in Table \ref{tab:chi2}).
This is also reflected in the slow light time lag panels in the top of Fig. \ref{fig:slowfastlightapp}.
Nevertheless, what has not changed is that a relatively large number of time lags passes the $\chi^2_\mathrm{o}$ criterion.
For the $\chi^2_\mathrm{f}$ constraint, however, we find that almost no time-lag passes, which mostly explains the somewhat diminished size in promising window size (grey horizontal bars) when compared to the fast light equivalent in Fig. \ref{fig:M0lightcurve+CC}.
This indicates that while there is a relatively good consistency when the time lag is compared to the observational data directly, it is not consistent with the proposed linear fit.
This is mostly because the slow light time lags display somewhat shallower slopes.
We note that W2 in Fig. \ref{fig:slowfastlightapp} corresponds almost exactly to W3 shown in Fig. \ref{fig:M0lightcurve+CC}.
While we find a similar $\chi^2_\mathrm{o} \approx 7$ for both the slow and fast time lags, we find that $\chi^2_\mathrm{f}$ differs significantly, which highlights the sensitivity of this particular diagnostic.
Nevertheless, if we focus on the more agnostic direct observational comparison ($\chi^2_\mathrm{o}$), then we find good agreement overall with the observations.

To conclude, we comment on why the slow light time lags can differ substantially from their fast light counterpart.
When employing the simplest interpretation, there are two main ways in which the light curves change when comparing fast with slow light results, namely: the timing between (light curve or image) features differs or the features themselves are different.
While the former point is largely determined by relative emission structure sizes and the associated light travel time differences, the latter point is directly associated with the radiative transfer calculations.
The preferred shallower time-lag slope does indicate a change in timing and from the light curves, we can already ascertain the more peaky nature of the slow light curves.
For the slow light approach, we would expect to find differences in regions where the plasma moves with or against the (integration) direction of the light rays, which either results in increased or reduced emission, respectively.
The higher the velocity of plasma, the greater this effect will be.
Another clear effect is related to the size of the emission structure, which is smaller for the higher frequencies indicating shorter light travel times through it.
Therefore, we would expect the differences at higher frequencies to be smaller than at low frequencies.
From this preliminary study, it becomes clear that the adaption of a slow light approach is able to significantly alter the light curves and underlying images at the evaluated wavelengths. Also, it does not necessarily result in better correspondence to the observational time lags (for the evaluated window).

\section{Discussion and conclusions} \label{sec:discconc}

We have demonstrated that flux eruption events and the emergence of flux tubes are consequential for the emission features at radio frequencies ($\nu = 19-47$ GHz).
Selected windows from our simulations are fully able to reproduce the characteristic observational time lags as presented in \citetalias{brinkerink15,brinkerink21}.
Interestingly, flux tube emission is prominently featured in the best-fitting time-lag windows, but that does not mean that the classical picture with an expanding and cooling jet sheath is no longer applicable. 
We advocate a more complex picture where the flux eruption, subsequent flux tube creation, and jet sheath emission structure are all affecting and perturbing one another, which is clearly seen in the most promising windows (in Fig. \ref{fig:grrtmultichrome}).
What happens to these features at larger distances still remains an open question.
As our simulations display relatively strong outflows extending several hundreds of \rg, we consider the exact origin of the emission in VLBI jet \citep{lobanov99,kim23} observations, which could perhaps be explained with a strong outflow (variability) starting at the central black hole rather than the (re)collimation shock picture.
However, we also note that our modeling only deals with the inner $500$ \rg and that these VLBI jet observations span thousands of \rg. 

A number of effects that are not within the scope of this work are likely to influence our results.
For the sake of improvement, we may consider the addition of non-thermal emission \citep{ozel00,chan09,chael17,davelaar18}, along with a GRMHD simulation focusing on better resolving the jet sheath \citep[as seen in, e.g.,][]{ripperda22} and more elaborate electron heating prescriptions \citep[see, e.g.,][]{howes10,rowan17,rowan19,kawazura19}; alternatively, even a more in-depth R$\beta$ study could be useful. 
Another possibility would be the inclusion of non-ideal currents in the vicinity of the jet-disk interface with a resistive GRMHD approach \citep{ripperda20,vos23}.
Additionally, polarized light curves \citep[see, e.g.,][]{moscibrodzka21,mahdi23,davelaar23} are likely to shed more light on the preferred magnetic field orientation -- if it diverges significantly from the commonly applied singular poloidal loop, as outlined in Sect. \ref{meth:grmhd}.

In this work, we have also undertaken an exploratory study to establish the sensitivity of our results to the fast-light assumption and found that it does indeed have a significant effect on the recovered time lags (see Sect. \ref{res:slow_light}).
At face value, the slow light time lags did not exhibit a better correspondence with the observational findings, which highlights the fact that the fast versus slow light paradigm is not intuitively understood, especially for a timing-sensitive study (as is the case here).   
Nevertheless, we have demonstrated that it is possible to explain the observed time lags originating from Sgr A$^\ast$ utilizing the modeling techniques outlined in this work. 

Lastly, we note that the time-lag slope of $A = 41 \pm 14$ cm/min \citepalias{brinkerink15,brinkerink21} is only recovered on relatively rare occasions (see Appendix \ref{app:bestwindows}) and seems to correspond to a particular configuration of the emission structure. Here, the flux tube starts its orbit on the front-right side of the main emission structure and recedes with respect to the observer over the evaluated time window. 
Nevertheless, as noted explicitly in \citetalias{brinkerink21}, observational time lags tend to span across a range of $20 \lesssim A \lesssim 40$ cm/min and a considerable part of the simulation time-lag distributions is consistent with this range as shown in Fig. \ref{fig:slopeapp}.
Our analysis favors a strong flux tube emission component to be present to explain the high observational time-lag slope of $A \approx 40$ cm/min.
Overall, it seems that the $a_\ast=0$ (\MZ) case outperforms the others two spin cases ($a_\ast=0.9375$ and $a_\ast=-0.9375$ for \MP and \MM, respectively) and has a more variable time-lag signature as shown in the confidence intervals in Fig. \ref{fig:slopeapp}.
This is likely tied to the comparative flux tube to base emission structure strength, where we expect the base emission structure of \MZ case to be less prominent due to the suppression of the BZ mechanism.   
In a future study, we aim to investigate the implications of our findings at higher frequencies and including polarization.

\section*{Acknowledgements}
We thank Monika Mo\'scibrodzka for the insightful discussions and comments that have improved the quality of the manuscript. 
J.V. acknowledges support from the Dutch Research Council (NWO) supercomputing grant No. 2021.013. J.D. is supported by a Joint Columbia University and Flatiron Institute Postdoctoral Fellowship. Research at the Flatiron Institute is supported by the Simons Foundation. J.D. acknowledge support from NSF AST-2108201.
During part of this project, funding for H.O. came from Radboud University Nijmegen through a Virtual Institute of Accretion (VIA) postdoctoral fellowship from the Netherlands Research School for Astronomy (NOVA).
This work is supported by the Center for Research and Development in Mathematics and Applications (CIDMA) through the
Portuguese Foundation for Science and Technology (FCT - Fundação para a Ciência e a Tecnologia), references
UIDB/04106/2020, UIDP/04106/2020. H.O. acknowledges support from the projects PTDC/FIS-AST/3041/2020, CERN/FIS-PAR/0024/2021 and 2022.04560.PTDC. This work has further been supported by the European Union’s Horizon 2020 research and innovation (RISE) programme H2020-MSCA-RISE-2017 Grant No. FunFiCO-777740 and by the European Horizon Europe staff exchange (SE) programme HORIZON-MSCA-2021-SE-01 Grant No. NewFunFiCO-10108625. H.O. is supported by the Individual CEEC program - 5th edition funded by the FCT.

\textit{Software:}
{\tt python} \citep{oliphant07python,vanrossum09python}, {\tt scipy} \citep{scipy20}, {\tt numpy} \citep{harris20numpy}, {\tt matplotlib} \citep{hunter07matplotlib}, {\tt RAPTOR} \citep{bronzwaer18,bronzwaer20}, {\tt BHAC} \citep{porth17,olivares19}.


\bibliographystyle{aa}
\bibliography{references}


\appendix

\section{Animations} \label{app:animations}

\begin{figure}[!h]
    \centering
    \includegraphics[width=\columnwidth]{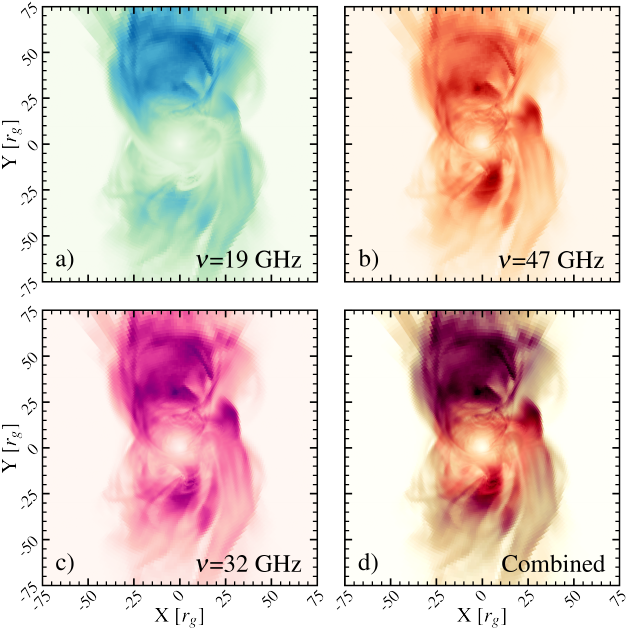}
    \caption{Schematic outlining the components of the multi-frequency graphics as shown in Fig. \ref{fig:grrtmultichrome}. The panels ($a$,$b$,$c$) display the flux density maps ($F_\nu$) at frequencies $\nu=19,32,47$ at $T=9230$ of the \MM, $i=30^\circ$ case, which is the same instance as shown in panels ($c$) and ($f$) of Fig. \ref{fig:3DGRMHD}. The image's RGBA codes are combined into the final result in the bottom right (panel $d$).
    }
    \label{fig:compositeschematic}
\end{figure}

To display the time lag in a visually intuitive manner, we combined the images at multiple frequencies into a single multi-frequency image, from which one can interpret regions of simultaneous emission or when a feature is predominantly present at a single frequency.
The colors per frequency are as follows: green-blue for $\nu=19$ GHz, pink-purple for $\nu=32$ GHz, orange-red for $\nu=47$ GHz that are combined into a red to dark-purple and black master color scheme.
All images saturate at a flux density ($F_\nu$) per pixel of $0.025$, which allows for interpreting which image feature (at which frequency) is most dominant.
As we modify the RGBA codes of the base figures to acquire the top figure, there is a potential loss of some physical interpretability, but the visual interpretation of the time lag becomes more intuitive.
The interpretation of the strongest emission features is naturally still robust.
An example of how the combination of base images occurs is shown in Fig. \ref{fig:compositeschematic}.
As mentioned, all $i=30^\circ$ simulations have accompanying animations that are available in this online repository\footnote{\href{https://doi.org/10.5281/zenodo.10046402}{https://doi.org/10.5281/zenodo.10046402}}.

\section{Accretion rate for all simulations} \label{app:accretion_rate}

Table \ref{tab:accretionrateapp} lists the acquired accretion rates as $\dot{M}_\mathrm{cgs} = \mathcal{M} \cdot \langle \dot{M}_\mathrm{sim} \rangle / t_\mathrm{g}$ as a function of the time-averaged simulation accretion rate ($\langle \dot{M}_\mathrm{sim} \rangle$) and the user-defined mass unit, $\mathcal{M,}$ as outlined previously in Sects. \ref{meth:grrt} and \ref{res:SED}.
The flux calibration procedure, which we outline in more detail here, is based on incrementally changing $\mathcal{M}$ based on the average flux acquired from a $100$ \rgc spaced light curve.
After updating $\mathcal{M}$, {\tt RAPTOR} is launched 900 times as the total length of the light curve is $9000$ \rgc. The $\mathcal{M}$ space is sampled by means of a binary search and typically converges on the desired flux density ($F_\nu = 2.5$ Jy at 230 GHz) in ten cycles or less.

\begin{table}
    \centering
    \caption{Accretion rates for all GRMHD simulations.}
    \begin{tabular}{ *4c } \toprule
        \emph{Model} & $i$ & $\mathcal{M}$ & $\dot{M}_\mathrm{cgs}$ \\  
        \emph{Name} & & [c.u.] & [$M_\odot$/yr] \\ \midrule
        {\tt MADP} & $10^\circ$ & $4.056742 \times 10^{17}$ & $5.522208 \times 10^{-9}$ \\  
                   & $30^\circ$ & $4.177095 \times 10^{17}$ & $5.686037 \times 10^{-9}$ \\  
                   & $50^\circ$ & $4.177095 \times 10^{17}$  &$5.686037 \times 10^{-9}$  \\ \midrule 
        {\tt MAD0} & $10^\circ$ & $9.684549 \times 10^{17}$ & $1.636448 \times 10^{-8}$  \\  
                   & $30^\circ$ & $9.024831 \times 10^{17}$  &$1.524972 \times 10^{-8}$  \\  
                   & $50^\circ$ & $8.126574 \times 10^{17}$ & $1.373189 \times 10^{-8}$ \\ \midrule 
        {\tt MADM} & $10^\circ$ & $7.225904 \times 10^{17}$ & $1.682900 \times 10^{-8}$ \\  
                   & $30^\circ$ & $6.943015 \times 10^{17}$ & $1.617016 \times 10^{-8}$ \\  
                   & $50^\circ$ & $6.378044 \times 10^{17}$ & $1.485435 \times 10^{-8}$ \\
    \bottomrule
    \end{tabular}
    \tablefoot{The average mass accretion rate values, $\dot{M}_\mathrm{cgs}$, are calculated following the steps outlined in Sect. \ref{res:SED}.
    }
    \label{tab:accretionrateapp}
\end{table} 

\section{Grid layout of the GRMHD simulations and flux tube temperature profiles} \label{app:grid}

\begin{figure*}
    \centering
    \includegraphics[width=\textwidth]{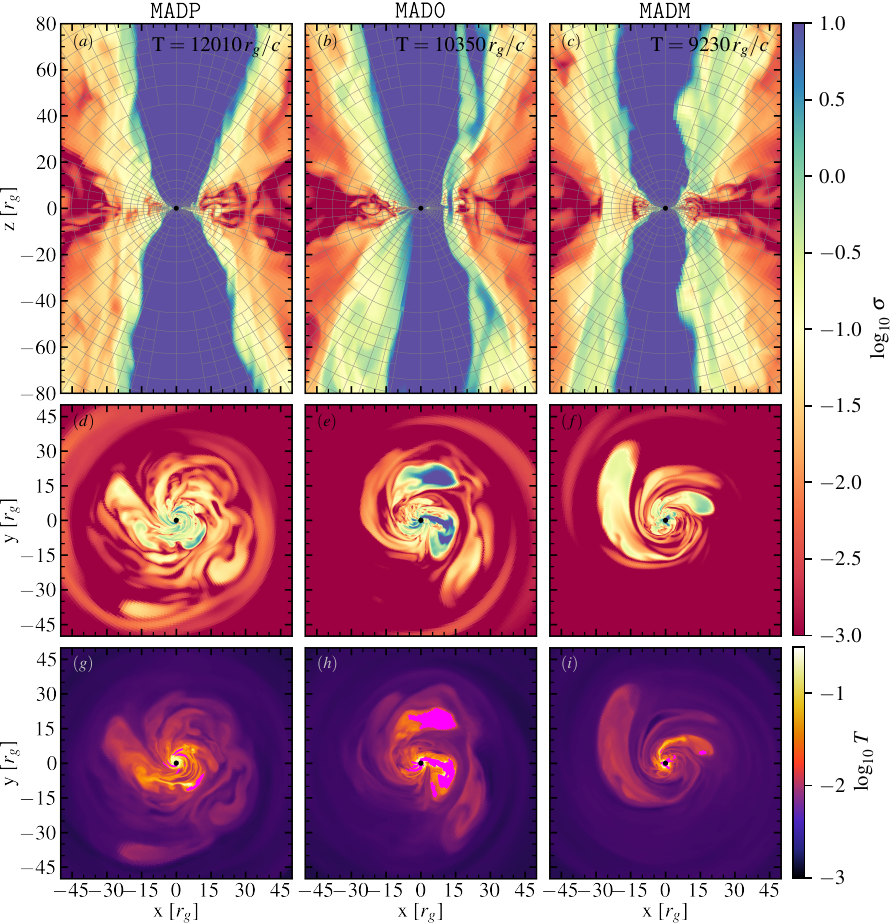}
    \caption{Flux tubes in the 3D GRMHD simulations. They visualized with meridial (panels $a$, $b$, $c$) and equatorial (panels $d$ till $i$) slices. Here, in addition to Fig. \ref{fig:3DGRMHD}, we display the (cold) magnetization $\sigma = B^2 / \rho$ (panels $a$ till $f$) with the AMR structure (in grey) and temperature $T=p/\rho$ (panels $g$ till $i$). Each AMR block contains $8 \times 8 \times 8$ cells. For intuitive comparison with the GRRT images, we display $T$ regions with $\sigma>1$ in magenta as the emissivity is set to zero there. As mentioned, the chosen slices correspond to promising time-lag windows and are listed in Table \ref{tab:chi2}. }
    \label{fig:gridtempapp}
\end{figure*}

Figure \ref{fig:gridtempapp} displays the AMR grid layout of our simulations, with a high refinement block covering the jet-disk boundary, and gives insight into the temperature profile of the flux tubes.
The high refinement blocks allow for capturing structural variability in the jet sheath better than is standardly applied.
As it is typically advantageous to keep the blocks at lowest level along the jet axis, we opted for the user-defined and intricate refinement scheme.
Nevertheless, to sufficiently resolve plasma-driven (i.e., Kelvin-Helmholtz instabilities; \citealt{sironi21}) or even current-driven (i.e., magnetic reconnection; \citealt{ripperda20,vos23}) instabilities one needs much higher resolution levels to resolve these properly.
Additionally, as discussed in Sect. \ref{res:qualitative}, we find that tail-end of the flux tube is both heated and over-dense which gives it a clear emission feature, while the very center of the tube (having low density) is typically excluded from the GRRT analysis (as $\sigma > 1$).

\section{Best-fit time windows} \label{app:bestwindows}

Table \ref{tab:chi2} lists the time-windows of the simulated time lags that fit best to the relation found in \citetalias{brinkerink15}. 
These windows are identified by means of $\chi^2_\mathrm{f}$ (related to the linear fit with slope $A=40\pm14$ cm/min) or $\chi^2_\mathrm{o}$ (direct comparison with the six observational time lags). 
The main identification criteria ($\chi^2_\mathrm{x}<7$, $\mathrm{x} \in \{\mathrm{o},\mathrm{f}\}$) are stringent, which ensure that we only identify the very best-fitting time-windows. We note that the reduced chi-squared, $\chi^2_\nu$, corresponding to the aforementioned criteria are $\chi^2_{\nu,\mathrm{o}} = 1.17$ and $\chi^2_{\nu,\mathrm{f}} = 0.58$, which clearly outlines that the $\chi^2_\mathrm{f}$ criterion is the main decider for the best-fitting windows. This also explains the main trends shown in the table, which we briefly discuss now.

Generally, from all evaluated cases, {\tt MAD0} is best able to recover the desired time lag, especially for inclination of $i=30^\circ$ and $i=50^\circ$. 
Overall, column (i) outlines that numerous light curve sections have passable correspondence to the observations, where the $i=10^\circ$ cases are slightly disfavored (especially for \MP and \MM). 
Column (ii) is where the clear preference arises for the \MZ case, as it seems that the other cases are less able to recover the needed slope which is also clearly seen for Fig. \ref{fig:slopeapp}. 
Columns (iii) and (iv) further accentuate the previously discussed point, but are nevertheless useful for selecting the very best time-lag windows. Nevertheless, we note that the $\chi^2_{\mathrm{o}} < 7$ criterion is leading as it is (most) free from any predefined assumption (i.e., linearity with a certain slope). From this result, we conclude that there are numerous windows that are consistent, but there are indications for a preferred medium inclination ($i=30^\circ/50^\circ$) and low BH-spin (\MZ). 

\begin{table*}
    \centering
    \caption{Best-fitting simulation windows.}
    \begin{tabular}{ l *8c } \toprule
        \emph{Model} & Incl. & LCCF win. & (i) ($\chi^2_\mathrm{o} < 8$) & (ii) ($\chi^2_\mathrm{f} < 8$) & (iii) ($\chi^2_\mathrm{o} < 8$) & (iv) ($\chi^2_\mathrm{o} < 7$) & Best \\  
        \emph{} & & [$r_g/c$] & & & \& ($\chi^2_\mathrm{f} < 25$) & \& ($\chi^2_\mathrm{f} < 7$) & window \\ \midrule
        {\tt MADP} & 10$^\circ$ & 100 & 121 [13.6\%] & 0 [0.0\%] & 0 [0.0\%] & 0 [0.0\%] & - \\
                   & 10$^\circ$ & 200 & 78 [8.9\%] & 0 [0.0\%] & 0 [0.0\%] & 0 [0.0\%] & - \\
                   & 10$^\circ$ & 300 & 116 [13.3\%] & 0 [0.0\%] & 0 [0.0\%] & 0 [0.0\%] & - \\
                   & 10$^\circ$ & 400 & 174 [20.2\%] & 0 [0.0\%] & 0 [0.0\%] & 0 [0.0\%] & - \\ \addlinespace
                   & 30$^\circ$ & 100 & 198 [22.2\%] & 0 [0.0\%] & 3 [0.3\%] & 0 [0.0\%] & - \\
                   & 30$^\circ$ & 200 & 199 [22.6\%] & 4 [0.5\%] & 20 [2.3\%] & 2 [0.2\%] & 11810, 13180 \\
                   & 30$^\circ$ & 300 & 224 [25.7\%] & 0 [0.0\%] & 2 [0.2\%] & 0 [0.0\%]  & -\\
                   & 30$^\circ$ & 400 & 245 [28.5\%] & 0 [0.0\%] & 0 [0.0\%] & 0 [0.0\%]  & -\\ \addlinespace
                   & 50$^\circ$ & 100 & 196 [22.0\%] & 0 [0.0\%] & 1 [0.1\%] & 0 [0.0\%]  & -\\
                   & 50$^\circ$ & 200 & 217 [24.6\%] & 2 [0.2\%] & 5 [0.6\%] & 1 [0.1\%]  & 12790\\
                   & 50$^\circ$ & 300 & 252 [28.9\%] & 1 [0.1\%] & 10 [1.1\%] & 0 [0.0\%] & -\\
                   & 50$^\circ$ & 400 & 288 [33.4\%] & 7 [0.8\%] & 17 [2.0\%] & 0 [0.0\%] & -\\ \midrule
        {\tt MAD0} & 10$^\circ$ & 100 & 180 [20.2\%] & 0 [0.0\%] & 4 [0.4\%] & 0 [0.0\%]  & -\\
                   & 10$^\circ$ & 200 & 191 [21.7\%] & 1 [0.1\%] & 10 [1.1\%] & 0 [0.0\%] & -\\
                   & 10$^\circ$ & 300 & 191 [21.9\%] & 7 [0.8\%] & 26 [3.0\%] & 2 [0.2\%] & 9910 - 9920\\
                   & 10$^\circ$ & 400 & 171 [19.9\%] & 19 [2.2\%] & 28 [3.3\%] & 3 [0.3\%] & 7920 - 7930, 10410\\ \addlinespace
                   & 30$^\circ$ & 100 & 197 [22.1\%] & 0 [0.0\%] & 6 [0.7\%] & 0 [0.0\%]  & -\\
                   & 30$^\circ$ & 200 & 184 [20.9\%] & 6 [0.7\%] & 20 [2.3\%] & 2 [0.2\%] & 7560, 9130\\
                   & 30$^\circ$ & 300 & 168 [19.3\%] & 14 [1.6\%] & 35 [4.0\%] & 6 [0.7\%] & 9910, 10330-10370\\
                   & 30$^\circ$ & 400 & 196 [22.8\%] & 31 [3.6\%] & 55 [6.4\%] & 8 [0.9\%] & 10310-10360, 11270-11280\\ \addlinespace
                   & 50$^\circ$ & 100 & 238 [26.7\%] & 0 [0.0\%] & 6 [0.7\%] & 0 [0.0\%] & -\\
                   & 50$^\circ$ & 200 & 222 [25.2\%] & 8 [0.9\%] & 30 [3.4\%] & 2 [0.2\%] & 10350-10360\\
                   & 50$^\circ$ & 300 & 235 [27.0\%] & 27 [3.1\%] & 47 [5.4\%] & 7 [0.8\%] & 9080-9090, 10350-10390\\
                   & 50$^\circ$ & 400 & 267 [31.0\%] & 43 [5.0\%] & 61 [7.1\%] & 3 [0.3\%] & 9860, 10370-10380\\ \midrule
        {\tt slow} & 30$^\circ$ & 100 & 49 [18.1\%] & 0 [0.0\%] & 0 [0.0\%] & 0 [0.0\%] & - \\
        {\tt MAD0} & 30$^\circ$ & 200 & 69 [27.5\%] & 0 [0.0\%] & 8 [3.2\%] & 0 [0.0\%] & - \\
                   & 30$^\circ$ & 300 & 78 [33.8\%] & 0 [0.0\%] & 1 [0.4\%] & 0 [0.0\%] & - \\ 
                   & 30$^\circ$ & 400 & 93 [44.1\%] & 0 [0.0\%] & 0 [0.0\%] & 0 [0.0\%] & - \\ \midrule
        {\tt MADM} & 10$^\circ$ & 100 & 74 [8.3\%] & 0 [0.0\%] & 1 [0.1\%] & 0 [0.0\%] & -\\
                   & 10$^\circ$ & 200 & 76 [8.6\%] & 0 [0.0\%] & 6 [0.7\%] & 0 [0.0\%] & -\\
                   & 10$^\circ$ & 300 & 85 [9.8\%] & 2 [0.2\%] & 16 [1.8\%] & 0 [0.0\%]& -\\
                   & 10$^\circ$ & 400 & 82 [9.5\%] & 5 [0.6\%] & 5 [0.6\%] & 0 [0.0\%] & -\\ \addlinespace
                   & 30$^\circ$ & 100 & 162 [18.2\%] & 0 [0.0\%] & 2 [0.2\%] & 0 [0.0\%]  & -\\
                   & 30$^\circ$ & 200 & 182 [20.7\%] & 1 [0.1\%] & 6 [0.7\%] & 0 [0.0\%]  & -\\
                   & 30$^\circ$ & 300 & 184 [21.1\%] & 3 [0.3\%] & 2 [0.2\%] & 0 [0.0\%]  & -\\
                   & 30$^\circ$ & 400 & 237 [27.5\%] & 0 [0.0\%] & 11 [1.3\%] & 0 [0.0\%] & -\\ \addlinespace
                   & 50$^\circ$ & 100 & 224 [25.1\%] & 0 [0.0\%] & 5 [0.6\%] & 0 [0.0\%]  & -\\
                   & 50$^\circ$ & 200 & 237 [26.9\%] & 1 [0.1\%] & 3 [0.3\%] & 0 [0.0\%]  & -\\
                   & 50$^\circ$ & 300 & 310 [35.6\%] & 4 [0.5\%] & 22 [2.5\%] & 2 [0.2\%] & 9100-9110\\
                   & 50$^\circ$ & 400 & 368 [42.7\%] & 11 [1.3\%] & 39 [4.5\%] & 6 [0.7\%] & 8990, 9010-9040, 14450\\
     \hline
    \end{tabular}
    \tablefoot{The total number of LCCF windows per light curve are 891, 881, 871, and 861 that correspond to a length of 100, 200, 300, and 400 \rgc, respectively. The "best window" column lists the starting time of the window in question. We note that the slow light case, {\tt slow MAD0}, has a shorter light curve length and is evaluated at a higher cadence of 5 \rgc; see Fig. \ref{fig:slowfastlightapp}. 
    }
    \label{tab:chi2}
\end{table*} 

\section{Different reference frequency} \label{app:different_reference_freq}

We have exclusively reported our results using the reference frequency of $\nu_1 = 19$ GHz.
Even though we typically find very consistent time lags when choosing a different reference frequency, it is still possible that the time-lag behavior changes.
Here, we show the differences in results when using a reference frequencies of $\nu_1 = 32$ GHz.
As mentioned before, only the slope of the relation is important for our analysis, as we would naturally get a different offset when changing the reference frequency.
In essence, we will therefore be repeating the analysis outlined in Sects. \ref{meth:linfit} and \ref{res:quantitative_time_lags} to estimate how much the choice in reference frequency affects the final time lag interpretation.

\begin{figure*}
    \centering
    \includegraphics[width=\textwidth]{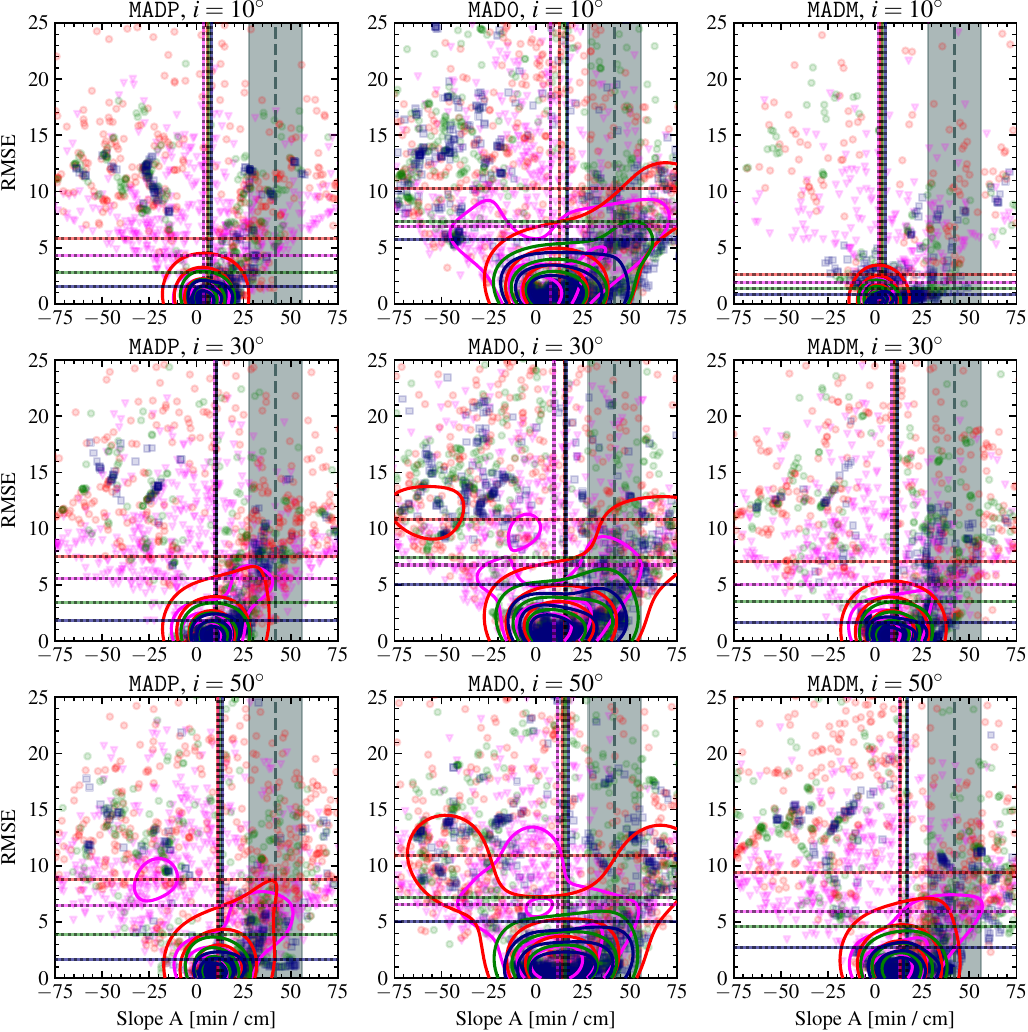}
    \caption{
    RMSE between the time lags acquired from the simulations and their linear fits for a reference frequency of $\nu_1 = 32$ GHz. 
    The points are color-coded according to the time-lag window length; 100 (\emph{magenta}), 200 (\emph{red}), 300 (\emph{green}), and 400 (\emph{blue}) \rgc.
    The color-dashed lines denote the mean slope of the best-fitting (RMSE < 5) time lags for the vertical lines, while the horizontal lines denote the mean RMSE of the entire population.
    The grey dashed line, with standard deviation area, corresponds to the linear fit to the VLA-only data from \citetalias{brinkerink15}, which is $A = 42 \pm 14$ cm/min. The contours denote the $25\%$, $50\%$, and $75\%$ levels of the normalized kernel density estimations for each of the window length results (in the corresponding colors). Figure \ref{fig:slopeapp} displays the results for a reference frequency of $\nu_1 = 19$ GHz, which was the base assumption throughout this work.
    }
    \label{fig:nu05slopeapp}
\end{figure*}

Based on the comparison of Fig. \ref{fig:nu05slopeapp} ($\nu_1 = 32$ GHz) and Fig. \ref{fig:slopeapp} ($\nu_1 = 19$ GHz), we conclude that there are differences, which are best pointed out by the density contours, between both time-lag distributions, but overall they are in good agreement.
More specifically, the variation in the distribution itself is lower (i.e., density contours are more confined) and the best-fitting linear slopes (for $\mathrm{RMSE} < 5$) are slightly lower than what was found for $\nu_1 = 19$ GHz.
Nevertheless, we still find a good number of time-lag slopes that lie within the shaded region. 
In our experience, when the reference light curve is smooth or devoid of small-scale features, it results in a more ``tolerant'' time-lag profile, which often contains competing features.
When we correlate the light curves of neighboring frequencies, we are able to acquire the clearest outcomes. 
Although the reference light of $\nu_1 = 19$ potentially results in a more variable time-lag profile, we find that the results and the interpretation are still robust, even when a different reference frequency is chosen.

\end{document}